%% file: main.tex
\documentclass[10pt, conference, letterpaper]{IEEEtran}
\IEEEoverridecommandlockouts
\usepackage{amsmath}
\usepackage{amsthm}
\usepackage{amsfonts}
\usepackage{graphicx}
\usepackage{xcolor}
\usepackage{subcaption}
\usepackage{comment}
\usepackage[ruled, vlined, linesnumbered]{algorithm2e}

\newcommand{\note}[1]{\textcolor{blue}{#1}}

\SetKw{Break}{break}

\newtheorem{lemma}{Lemma}
\newtheorem{theorem}{Theorem}

\def\eg{\textit{e.g.}}
\def\ie{\textit{i.e.}}

\newcommand{\yxbao}[1]{{\textcolor{green}{Yixin: #1}}}
\newcommand{\revise}[1]{{#1}}

\begin{document}
\title{Optimizing Task Placement and Online Scheduling for Distributed GNN Training Acceleration
\thanks{This work was supported in part by grants from Hong Kong RGC under the contracts HKU 17204619, 17208920 and 17207621.}
}
\author{\IEEEauthorblockN{Ziyue Luo\IEEEauthorrefmark{1}, Yixin Bao\IEEEauthorrefmark{1}, Chuan Wu\IEEEauthorrefmark{1}} 
	\IEEEauthorblockA{\IEEEauthorrefmark{1}Department of Computer Science, The University of Hong Kong, Email: \{zyluo, yxbao, cwu\}@cs.hku.hk}}
\maketitle

\begin{abstract}
Training Graph Neural Networks (GNN) on large graphs is resource-intensive and time-consuming, mainly due to the large graph data that cannot be fit into the memory of a single machine, but have to be fetched from distributed graph storage and processed on the go. Unlike distributed deep neural network (DNN) training, the bottleneck in distributed GNN training lies largely in large graph data transmission for constructing mini-batches of training samples. Existing solutions often advocate data-computation colocation, and do not work well with limited resources where the colocation is infeasible. The potentials of strategical task placement and optimal scheduling of data transmission and task execution have not been well explored. This paper designs an efficient algorithm framework for task placement and execution scheduling of distributed GNN training, to better resource utilization, improve execution pipelining, and expediting training completion. Our framework consists of two modules: (i) an online scheduling algorithm that schedules the execution of training tasks, and the data transmission plan;  and (ii) an exploratory task placement scheme that decides the placement of each training task. We conduct thorough theoretical analysis, testbed experiments and simulation studies, and observe up to 67\% training speed-up with our algorithm as compared to representative baselines.
\end{abstract}

\input{sec/intro}

\input{sec/background}
\input{sec/related_work}
\input{sec/model}
\input{sec/alg_flow}
\input{sec/alg_placement}

\input{sec/alg_complete}
\input{sec/eval}

\input{sec/conclusion}
\input{sec/appendix}

\newpage
\bibliographystyle{IEEEtran}
\bibliography{./ref}

\end{document}

%% file: sec/intro.tex
\section{Introduction}

Graph neural networks (GNN) \cite{kipf2016semi}\cite{hamilton2017inductive} generalize deep neural networks (DNN) to learning from graph-structured data and have been exploited in various domains, \eg, computer networking~\cite{rusek2020routenet}, social and biological network analysis~\cite{li2019encoding}\cite{duvenaud2015convolutional}. GNNs learn high-level graph representations (aka embeddings) by aggregating information from the neighborhood of nodes in a graph, and have shown their superiority in various tasks including node classification \cite{velivckovic2017graph}, graph classification \cite{ErricaPBM20} and link prediction \cite{li2020type}.

As compared to traditional graph analysis models~\cite{frasconi1998general}\cite{cao2015grarep}, GNNs can capture more complicated features of nodes/edges of large graphs with millions of nodes and billions of edges (\eg, Amazon Product Co-purchasing Network~\cite{hu2020open}, Microsoft Academic Graph~\cite{sinha2015overview}). 
However, training GNNs on large graphs is very resource-intensive and time-consuming. 
The large graph sizes often exceed the memory and computation capacities of a single device (\eg, GPU) or physical machine, yielding distributed GNN training using 
multiple devices and machines as the solution. While full-graph training by loading the entire graph into device memory is often infeasible~\cite{kipf2016semi}, a common practice of distributed GNN training is to do subgraph sampling
\cite{zeng2019graphsaint}\cite{chen2018fastgcn} and mini-batch training at each device: samplers select a set of training nodes in the graph, retrieve from graph stores features of (a subset of) several-hop neighbor nodes of each training node 
 to form subgraphs, construct mini-batches with the subgraphs and feed them into workers for training. 

A few distributed GNN training frameworks have recently been proposed, \eg,~distributed DGL~\cite{zheng2020distdgl}, 
Dorylus~\cite{thorpe2021dorylus}. 
It has been observed that frequent, large graph data transfers exist in distributed GNN training, as mini-batch sampling is carried out in each training iteration, which involves retrieval of subgraphs 
commonly consisting of hundreds of graph nodes each. 
Graph data transfer often consumes the majority of time during GNN training (up to 80\% of overall training time \cite{zheng2020distdgl}\cite{gandhi2021p3}) and renders the performance bottleneck of GNN training, which is different from the common bottlenecks of computation or gradient/parameter communication in DNN training.   
 Careful design to alleviate the graph data transfer overhead is hence the key for distributed GNN training acceleration.   

A few efforts have been devoted to 
minimizing the graph data transfers in distributed GNN training, through static caching~\cite{lin2020pagraph}, min-edge-cut graph partition~\cite{karypis1998fast}, and
 data-computation co-location~\cite{zheng2020distdgl}. Even with these 
schemes, 
large data transfers between samplers and graph stores may still exist; data-computation co-location may not always be applicable when resource availability varies across machines.
On the other hand, 
strategical task placement, data flow and task execution scheduling to improve resource utilization and execution parallelization, have not been well explored, which can be good complements to the traffic-minimizing schemes for distributed GNN training acceleration.

We focus on optimized planning of distributed GNN training, involving effective placements of training tasks (samplers, workers and parameter servers), near-optimal execution scheduling of the tasks, and data flow transfers. Unique challenges exist in distributed GNN training planning:

\textit{First}, existing designs largely advocate co-locating a worker with its samplers on the same physical machine, which is only applicable if the computational resources on the machine allow. 
In a practical machine learning (ML) cluster where resource availability differs across machines, it is non-trivial to plan task placements to minimize data transfer traffic and maximize resource utilization.

\textit{Next}, optimal scheduling of data transfers and task execution in a distributed GNN training job is complex, falling in the category of strongly NP-hard multi-stage coflow scheduling problems \cite{tian2018scheduling}.
Further, the data transfer volume between graph stores and samplers varies 
according to the graph nodes and their neighbors 
sampled in each training iteration~\cite{zeng2019graphsaint}\cite{chen2018fastgcn} and their storage locations, 
rendering the scheduling problem an online nature and calling for efficient online algorithm design.

Tackling the challenges, we design an algorithm framework for distributed GNN training planning, comprising two modules: 1) an online scheduling algorithm to strategically set execution time of training tasks and transfer rates of data flows; and 2) an exploratory task placement scheme that decides the placement of each task among available machines. Our goal is to maximize task parallelization while respecting various dependencies, and hence minimize the overall training time of a given GNN model. Our main techniques and contributions are summarized as follows:

$\triangleright$ 
Given task placements, we formulate the task and flow scheduling problem for distributed GNN training as an online optimization problem. We design an online scheduling algorithm by effectively overlapping task computation with graph data communication, and adaptively balancing the flow transmission rates among parallel flows into (from) the same machine, to eliminate negative impact of potential communication bottlenecks on the training time.
We rigorously analyze the online algorithm and identify a competitive ratio on the training makespan, which is decided by the maximum number of incoming or outgoing flows at any machine in one iteration.

$\triangleright$ Next, we propose an exploratory task placement scheme based on the Markov Chain Monte Carlo (MCMC) framework~\cite{geyer1992practical}. 
We start by efficient construction of an initial feasible placement in polynomial time. 
We then introduce a resource violation tolerance factor to encourage full exploration among feasible placements in the solution space.
A carefully designed cost function of the placements, defined on the expected training makespan and resource feasibility, guides our search process to the best feasible placement of tasks in arbitrary (heterogeneous) environments, to achieve the minimal expected training time in conjunction with our online scheduling algorithm.

$\triangleright$ We implement our design atop DGL~\cite{wang2019deep}, 
and conduct thorough testbed experiments and trace-driven simulations. Testbed experiments show that our design achieves significantly lower GNN training time as compared to DistDGL~\cite{zheng2020distdgl} 
(31.75\% on ogbn-products dataset~\cite{hu2020open} 
and 22.95\% on Reddit dataset~\cite{hamilton2017inductive})  
with more efficient network bandwidth utilization.
Simulation studies further prove that our design accelerates training up to 67\% compared to representative 
baselines under more diversified training settings, by exploiting strategical task placements to minimize the overall data traffic and maximize the utilization of heterogeneous network bandwidths, maximally overlapping communication with computation, and efficiently scheduling data traffic despite the varying data volumes.

%% file: sec/background.tex
\section{Background and Related Work}
\label{sec::background}

\subsection{GNN Training}

\begin{figure}[!ht]
  \centering
  \includegraphics[width=0.4\textwidth]{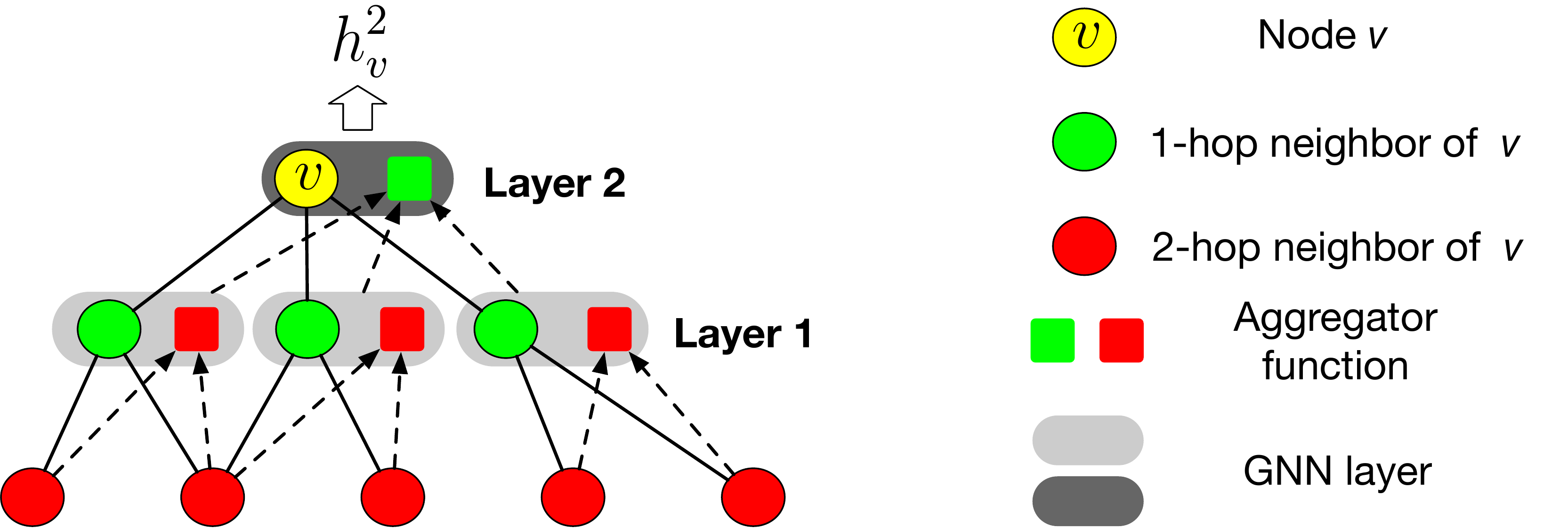}
  \caption{An example of embedding calculation of node $v$ with a 2-layer GNN: embeddings of $v$'s 1-hop neighbors ($l=1$) are computed using $h^l_v = \sigma(h_v^{l-1}, \text{\small AGGREGATE}_{v' \in v\text{'s neighbors}} f(h_v^{l-1}, h_{v'}^{l-1}, e_{v, v'}))$, and then aggregated to derive $v$'s embedding, $h^2_v$, using the formula with $l=2$. $\sigma(\cdot)$ and $f(\cdot)$ are trainable parameterized functions, $e_{v, v'}$ is the edge between $v$ and $v'$, and {\small AGGREGATE} is an aggregator function (\eg,~mean, min, max). 
  }
  \label{fig:gnn_training}
  \vspace{-7mm}
\end{figure}

GNNs learn effective graph 
embeddings by iteratively aggregating neighborhood features (Fig.~\ref{fig:gnn_training})~\cite{gilmer2017neural}\cite{cai2021rethinking}. 
The derived embeddings can be further processed (\eg,~using DNN layer, softmax operation), to produce decisions for downstream tasks (\eg, node classification, link prediction). 

To construct a mini-batch for GNN training, a set of training nodes are sampled from the input graph, and their $L$-hop neighbors are used for embedding generation by a $L$-layer GNN. 
Using features of all $L$-hop neighbors of the selected training nodes may lead to GPU/CPU memory overflow or high computation complexity. A common practice is to recursively sample neighbors of each training node with a sampling algorithm (\eg,~\cite{zeng2019graphsaint}\cite{chen2018fastgcn}), 
and a sub-graph is formed among the training node and its sampled $L$-hop neighbors. Each sub-graph with its features renders one sample in the mini-batch.

Using mini-batches of graph samples, GNN training is similar to DNN training: forward propagation is carried out to compute a loss, and then backward propagation to derive gradients of the GNN model parameters based on the loss, using an optimization algorithm (\eg,~stochastic gradient descent); a gradient update operation follows, which involves gradient aggregation among workers in distributed training and application of updated parameters to the GNN model.

%% file: sec/related_work.tex

\vspace{-2mm}
\subsection{Distributed GNN Training Systems}
Deep Graph Library (DGL)~\cite{wang2019deep} is a package built for easy implementation of GNN models on top of DL frameworks (\eg, PyTorch \cite{adam2019pytorch}, MXNet \cite{chen2015mxnet}). The recent release of DGL supports distributed GNN training on relatively large graphs. It uses random sampling, collocates one worker with one graph store, and does not pipeline GNN training across iterations, leaving a large room for further performance improvement. Euler~\cite{euler} is integrated with TensorFlow \cite{abadi2016tensorflow} for GNN training, which partitions a large graph in a round-robin manner 
and splits feature retrieving requests to allow concurrent transmissions; large data transfers still exist due to its locality-oblivious graph partition.
AliGraph~\cite{zhao2019aligraph} adopts distributed graph storage, optimized sampling operators and runtime to efficiently support GNNs. PyTorch Geometric~\cite{fey2019fast} is a deep learning library on irregularly structured input data such as graphs, supporting multi-GPU training on a single machine only.  
Dorylus~\cite{thorpe2021dorylus} distributes GNN training over serverless cloud function threads on CPU servers, requiring specialized functions provided by AWS~\cite{lambda}. 
Large data traffic exists in these systems, and careful transfer scheduling and task deployment can enhance them for training time minimization.

\subsection{Distributed 
Training Acceleration}

NeuGraph~\cite{ma2019neugraph} and PaGraph~\cite{lin2020pagraph}, which train GNN models on a single machine, adopt full-graph training by loading entire graphs into GPU memory, and are hence only feasible for training over small graphs. 
Considering multi-server clusters, ROC~\cite{jia2020improving} splits the input graph over multiple GPUs or machines to achieve workload balance, and adopts a memory management scheme to reduce CPU-GPU data transfer.
DistDGL~\cite{zheng2020distdgl} 
alleviates network transfer in distributed GNN training by co-locating each worker with its samplers on the same server, and partitioning the input graph with a minimum edge cut method. 
Further, various graph partition, sampling and caching methods have been proposed for enhancing distributed GNN training~\cite{hamilton2017inductive}\cite{wang2021flexgraph}\cite{chiang2019cluster}. These studies focus on minimizing data transfer volumes across devices/machines. 
Optimization of task placement and execution scheduling is orthogonal to the existing efforts, and our solution can complement them to fully accelerate distributed GNN training. 
DGCL~\cite{cai2021dgcl} is a recently proposed communication library for distributed GNN training, which decides data routing strategy for every graph node to the requiring worker(s), considering the detailed interconnection topology among workers. 
Its detailed communication plan is re-computed before every training epoch, which may incur substantial overhead for large graphs. Our design performs efficient, polynomial-time online scheduling on both task execution and data flows between tasks, effectively reducing the overall training time.

Task placement, computation and communication scheduling have been studied for DNN training on non-graph data \cite{zhang2017poseidon}\cite{Jayarajan2019p3}\cite{shi2021mg}. The communication scheduling deals with arranging transmission time and order of gradient/parameter tensors for parameter synchronization \cite{Jayarajan2019p3}\cite{shi2021mg}. Placement studies focus on worker placement to minimize interference~\cite{bao2019deep} instead of proximity to data,
and DNN operator placement to achieve model parallelism \cite{mirhoseini2017device}. Computation scheduling deals with fine-grained operator execution ordering, in case of model- or pipeline-parallel DNN training~\cite{wang2020geryon}\cite{park2020hetpipe}\cite{yi2020optimizing}. 
\revise{Compared to distributed DNN training, GNNs are largely trained with data parallelism, incurring large graph data communication that blocks the computation and occupies a majority of the training time (up to 80\%~\cite{gandhi2021p3}).}
Instead of operator-level placement and scheduling of a GNN model, 
we study placement of tasks (samplers, workers and parameter servers), overlap both graph data transfer and tensor communication with computation (the graph data traffic is magnitudes larger than tensor transfers), and pipeline mini-batch training across training iterations, which are all dedicated for GNN training acceleration.

%% file: sec/model.tex
\vspace{-2mm}
\section{Problem Model}
\label{sec::sys_model}
\vspace{-1mm}

\subsection{Distributed GNN Training System}

We train a GNN model (with $L$ embedding layers) in a cluster of $M$ physical machines. Partitions of a large graph used for GNN training are stored on the $M$ machines. Each machine $m \in [M]$\footnote{$[X]$ denotes set $\{1,2,\ldots, X\}$} is equipped with $R$ types of computational resources (\eg, GPU, CPU and memory), 
with type-$r$ resource available at the amount of $C^r_m$. Let $B^m_{in}$ ($B^m_{out}$) represent the available incoming (outgoing) NIC bandwidth on machine $m$.

There are four types of tasks in our GNN training job: 
(1) {\em Graph store server:} Each machine hosts a graph store server, to maintain one graph partition (including graph structure and node/edge features).
(2) {\em Sampler:} Each sampler selects training nodes, 
retrieves sampled node/edge features from graph store servers and forms sub-graphs. 
(3) {\em Worker:} Each worker carries out forward and backward computation, pushes gradients to and pulls parameters from parameter servers for parameter synchronization. A worker is typically associated with one or multiple samplers, which supply mini-batches dedicatedly to the worker. 
(4) {\em Parameter server (PS)}: PSs aggregate gradients from all workers, update the GNN model parameters and distribute updated parameters to all workers. 

We use $J_{g}, J_{s}, J_{w}$ and $J_{ps}$ to represent the sets of graph store servers, samplers, workers and PSs, respectively, in the training job. We suppose the number of each type of tasks is specified by the ML developer: the number of graph stores is $M$ (as each machine hosts exactly one graph partition), 
the number of workers can be larger or smaller than $M$ (considering a machine may host multiple GPUs and CPUs, and a worker typically consumes one GPU or CPU), the number of samplers to serve each worker is usually fixed (\eg, 2 samplers per worker). Let $J=J_{g}\cup J_{s}\cup J_{w} \cup J_{ps}$ denote the set of all tasks. 
Each task $j\in J$ occupies a $w_j^r$ amount of type-$r$ resource, $\forall r \in [R]$. For example, graph store servers, samplers and PSs are commonly run on CPUs, while workers can run on GPUs~\cite{lin2020pagraph} or CPUs~\cite{zheng2020distdgl}, and consume the respective memory. 
Tasks of the same kind (\eg, all samplers) occupy the same amount of resources.
Let $p_j$ denote the execution time of task $j$ in each iteration.



\begin{figure}[!t]
  \centering
  \includegraphics[width=0.47\textwidth]{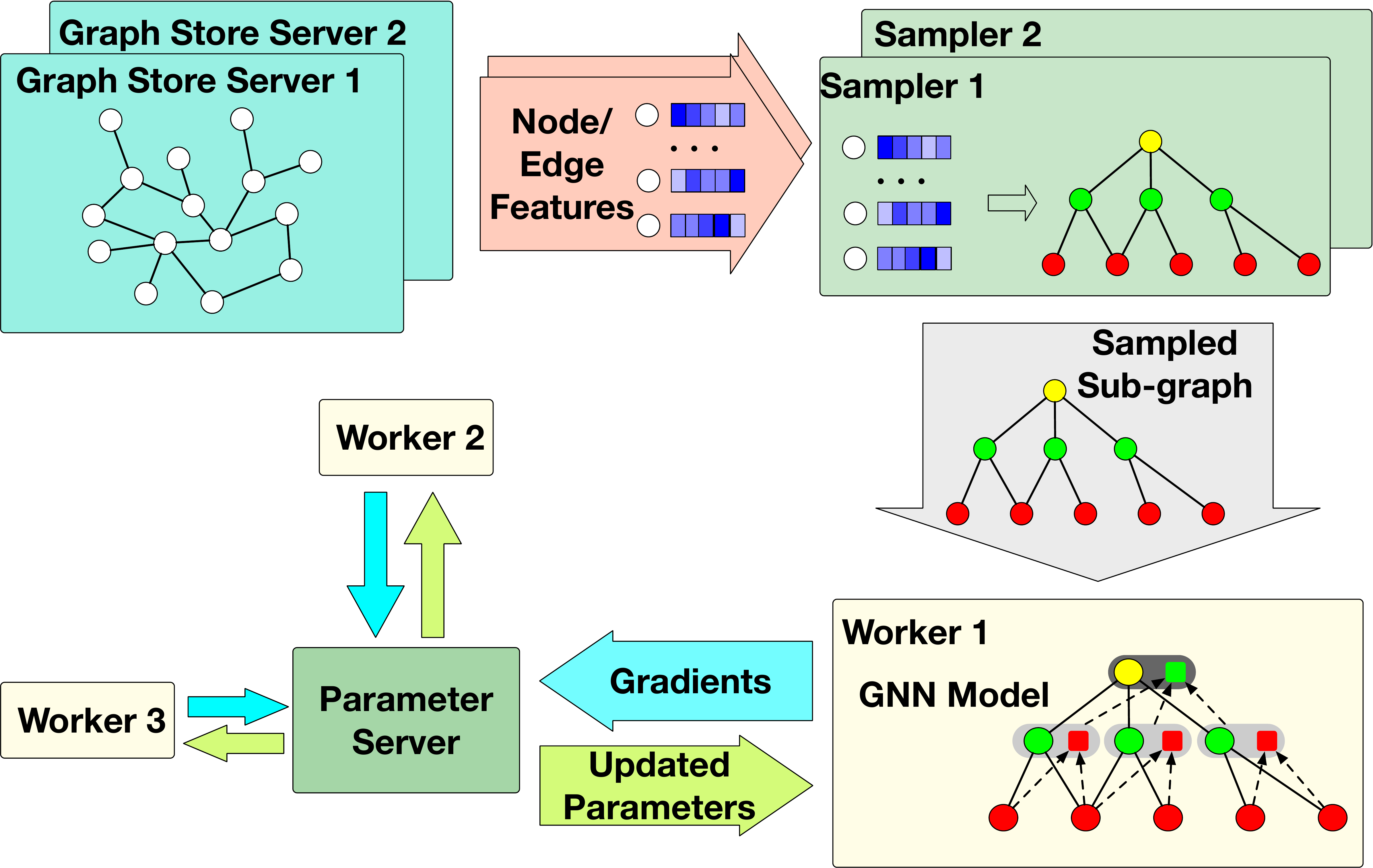}
  \caption{Distributed GNN training workflow}
  \label{fig:gnn_workflow}
  \vspace{-7mm}
\end{figure}

In a training iteration, each sampler selects a number of training nodes from the input graph and signals the graph store servers to acquire neighbor information. Upon requests from a sampler, a graph store server samples among $L$-hop neighbors of the training nodes that it hosts (using a given sampling algorithm), and sends the node/edge features back to the sampler. The sampler 
then sends sub-graph samples to its associated worker, which form a mini-batch from samples supplied by its sampler(s), for forward and backward computation. Computed gradients are sent from workers to the PSs and then updated parameters are dispatched from PSs to workers. The workflow 
is illustrated in Fig.~\ref{fig:gnn_workflow}.

\begin{figure*}[!t]
\centering
    \begin{subfigure}{0.6\columnwidth}
      \centerline{\includegraphics[width=1\columnwidth]{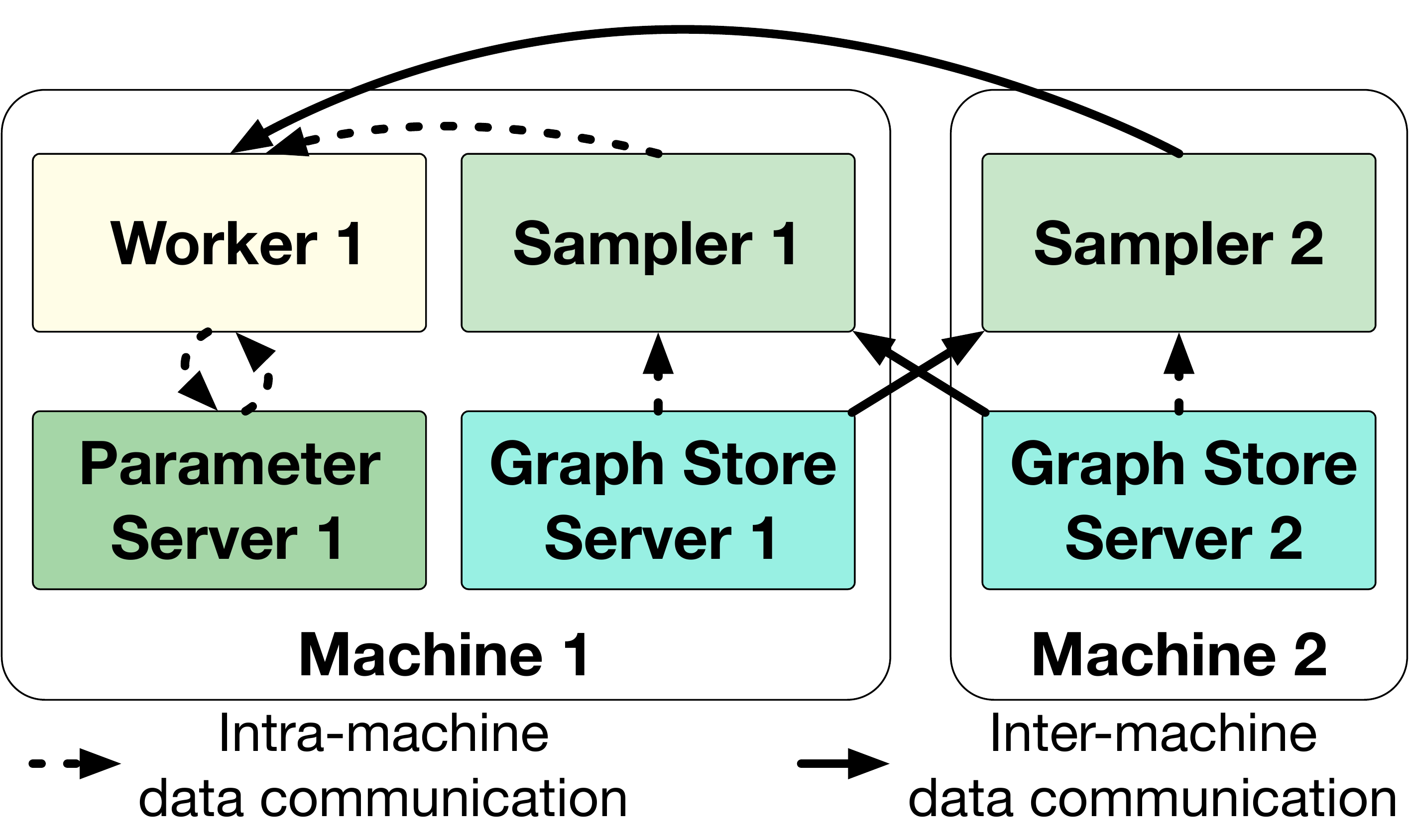}}
      \caption{Task placement}
    \end{subfigure}
    \begin{subfigure}{1.4\columnwidth}
      \centerline{\includegraphics[width=1\columnwidth]{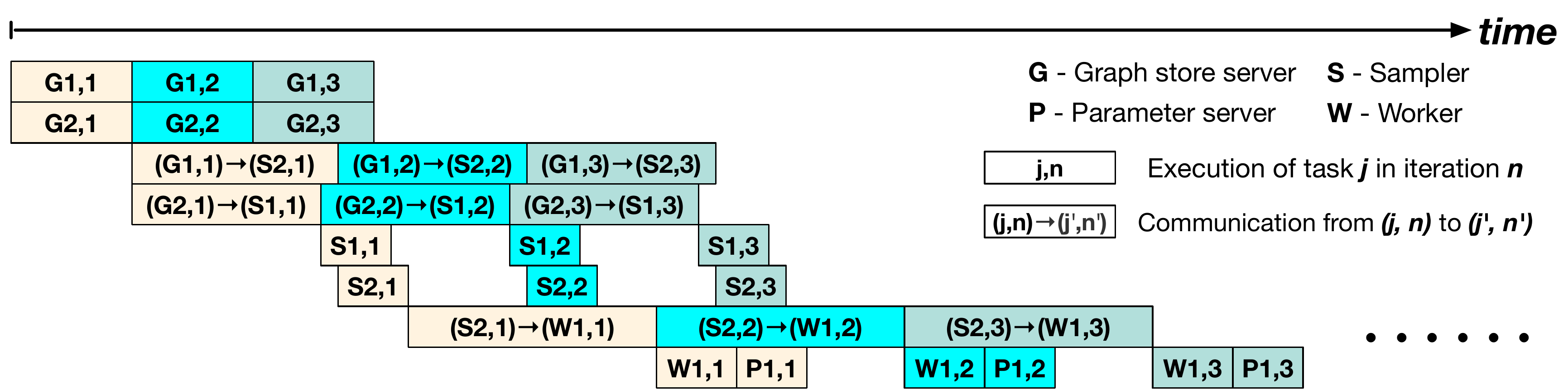}}
      \caption{Execution and flow schedule}
    \end{subfigure}
    \caption{A distributed GNN training planning example: a job with 2 graph store servers, 1 worker with 2 samplers, and 1 PS.}
    \label{fig:gnn_planning}
    \vspace{-6mm}
\end{figure*}

\subsection{Problem Formulation}
\vspace{-1mm}

We target overall training time minimization in our distributed GNN training job. Our design space includes two subproblems. 

\subsubsection{\bf Task Placement} We decide placements of all tasks in the GNN training job on the machines, to maximize task parallelization and minimize communication traffic. We use binary variable $y^m_j$ to indicate task placement: $y^m_j$ equals $1$ if task $j$ is deployed on machine $m$, and $0$, otherwise. The placement constraints are: 

\vspace{-5mm}
{\small
\begin{eqnarray}
\sum\limits_{m\in [M]}y^m_j= 1, \forall j\in J\label{eqn:place_1}\\
\sum\limits_{j\in J}w^r_jy^m_j \leq C^r_m, \forall m\in [M], r\in[R]\label{eqn:device_capacity}\\
y^{m}_j= 1, \forall j\in J_g \label{eqn:place_2}, j \text{~is placed on machine~} m\\
y_j^m\in \{0,1\}, \forall j\in J, m\in [M]
\end{eqnarray}
}
\vspace{-6mm}

\noindent Constraints in (\ref{eqn:place_1}) ensure that every task is placed on one and only one machine. (\ref{eqn:device_capacity}) are resource capacity constraints on the machines. (\ref{eqn:place_2}) specifies the given placements of graph store servers on machines. Fig.~\ref{fig:gnn_planning}(a) shows an example task placement of a GNN training job on two machines.


\subsubsection{\bf Online Execution and Flow Scheduling} Suppose it takes $N$ iterations for the GNN model training to converge. Given task placements, we decide the start time of each task and transmission schedules of sampled data and tensor flows, in each training iteration. 
Let binary variable $x_{j,n}^t$ indicate the start time of task $j$ in iteration $n$: $x_{j,n}^t$ is $1$ if task $j$ in iteration $n$ starts at time $t$, and $0$, otherwise. We use $k_{(j,n)\rightarrow (j',n')}^t$ to denote the amount of traffic sent from task $j$ of iteration $n$ to task $j'$ of iteration $n'$ at time $t$, including the following cases: sampled graph data from a graph store server to a sampler or from a sampler to a worker in the same iteration, gradients from a worker to a PS, or parameters updated at a PS ($j$) in iteration-$n$ training to a worker ($j'$) for iteration-$(n+1)$ training ($n'=n+1$).

The execution schedule should respect execution dependencies among tasks and flows, 
as follows:
\vspace{-5mm}

{\small
\begin{eqnarray}
x_{j, 1}^1 = 1, \forall j\in J_g\label{eqn:init_graph}\\
\sum\limits_{t\in[T]}x_{j, n}^t = 1, \forall j\in J, n\in[N]\label{eqn:time_restriction}\\
\min\{t|k_{(j,n)\rightarrow (j',n')}^{t}>0, t\in[T]\} \geq \sum\limits_{t\in[T]}tx^t_{j,n} + p_j, \forall j\in J, \nonumber\\
n\in[N], (j',n')\in succ(j,n), \text{$j$ and $j'$ are on different servers}\label{eqn:flow_start}\\
\max\{t|k_{(j,n)\rightarrow (j',n')}^{t}>0, t\in[T]\} < \sum\limits_{t\in[T]}tx^t_{j',n'}, \forall j\in J,\nonumber\\
 n\in[N],	(j',n')\in succ(j,n), \text{$j$ and $j'$ are on different servers} \label{eqn:flow_end}\\
 \sum\limits_{t \in[T]}tx_{j,n}^{t} + p_j \leq \sum\limits_{t\in[T]}tx_{j',n'}^{t}, 
	\forall j\in J, n\in[N], \nonumber\\
(j',n')\in succ(j,n), \text{$j$ and $j'$ are on the same server}\label{eqn:comp_same_server}
\end{eqnarray}
}
\vspace{-5mm}

We ignore the training node selection time at a sampler, and message passing from a sampler to a graph store server for graph data requests, as the traffic volume is negligible. Constraint~(\ref{eqn:init_graph}) indicates that graph store servers run first to sample neighbors. (\ref{eqn:time_restriction}) ensures that each task in each training iteration 
is scheduled once. Here $T$ is a potentially large time span in which our GNN training converges. 

Among tasks and flows, there are the following execution dependencies: (i) a sampler can start after receiving data from all graph store servers in each iteration;
(ii) in iteration $n$, a worker can start after receiving a mini-batch of graph data 
from its samplers and model parameters updated in iteration $n-1$; 
(iii) a PS can start after receiving gradients from all workers, computed in this iteration.
We call $(j', n')$ a successor of $(j, n)$ if task $j'$ in iteration $n'$ can only start after receiving data from task $j$ in iteration $n$, and $succ(j,n)$ denotes the set of all successors of $(j,n)$. Constraint (\ref{eqn:flow_start}) specifies that transmission from $(j,n)$ to its successor $(j',n')$ starts after $(j,n)$ is done. 
 (\ref{eqn:flow_end}) ensures that task $j'$ in iteration $n'$ does not start before the transfer from $(j,n)$ to $(j', n')$ is completed, if tasks $j$ and $j'$ do not reside on the same machine. We ignore data passing time between tasks on the same machine, but specify execution dependency among those tasks in~(\ref{eqn:comp_same_server}).

Across training iterations, we require that task $j$ in iteration $n+1$ can only start after task $j$'s execution in iteration $n$ has been done (\eg, a sampler prepares training data for iteration $n$ before those for iteration $n+1$), and data transfer $(j,n+1)\rightarrow (j',n'+1)$ cannot start before transmission $(j,n)\rightarrow (j',n')$ has been completed. These inter-iteration dependencies are formulated as in (\ref{eqn:iter_component}) and (\ref{eqn:flow_component}): 

\vspace{-3mm}
{\small
\begin{eqnarray}
	\sum\limits_{t\in[T]}tx_{j, n}^{t} + p_j \leq \sum\limits_{t\in[T]}tx_{j,n+1}^{t}, 
	\forall j\in J, n\in[N-1]\label{eqn:iter_component}\\
	\max\{t|k^t_{(j,n)\rightarrow (j',n')}>0, t\in[T]\} < \nonumber\\
	\min\{t|k^t_{(j,n+1)\rightarrow (j',n'+1)}>0, t\in[T]\}, \forall j\in J, n\in[N-1], \nonumber\\
	(j',n')\in succ(j,n),\text{$j$ and $j'$ are placed on different servers}\label{eqn:flow_component}
\end{eqnarray}
}
\vspace{-5mm}

Further, the following constraint specifies the total traffic transmitted from task $j$ in iteration $n$ to task $j'$ in iteration $n'$, as denoted by $d_{(j,n)\rightarrow (j',n')}$. The traffic volume is decided according to whether it is graph data transfer from a graph store server to a sampler or from a sampler to a worker (decided by the graph sampling algorithm in use), or gradient/parameter tensor transfer between a worker and a PS (depending on the GNN model size).

\vspace{-4mm}
{\small
\begin{eqnarray}
	\sum\limits_{t\in[T]}k^t_{(j,n)\rightarrow (j',n')} = d_{(j,n)\rightarrow (j',n')}, 
	\forall j\in J, n\in[N], \nonumber\\
	(j',n')\in succ(j,n),\text{$j$ and $j'$ are placed on different servers} 
\end{eqnarray}
}
\vspace{-4mm}

\noindent The total incoming (outgoing) traffic at machine $m$ should not exceed its available bandwidth at each time $t$: 

\vspace{-3mm}
{\small
\begin{eqnarray}
	\sum\limits_{n\in [N]}\sum\limits_{j\in J: y^m_j = 1}\sum\limits_{(j',n')\in succ(j,n):y^m_{j'}=0}k^t_{(j,n)\rightarrow (j',n')} \leq B^m_{out}, \nonumber\\
	\forall m\in[M], t\in[T]\\
	\sum\limits_{n\in [N]}\sum\limits_{j\in J: y^m_j = 0}\sum\limits_{(j',n')\in succ(j,n):y^m_{j'} = 1}k_{(j,n)\rightarrow (j',n')}^t \leq B^m_{in}, \nonumber\\
	\forall m\in[M], t\in[T]\label{eqn:band_in}
\end{eqnarray}
}
\vspace{-4mm}

We aim at minimizing the makespan of all $N$ iterations of GNN training, which is computed as $\max_{t \in {T}, j\in J}\{tx^t_{j, N}+p_j\}$. Given task placements $\{y_j^m\}$, the execution and flow scheduling problem is formulated as:

\vspace{-3mm}
{\small
\begin{equation}
	\label{eqn:flow_schedule_problem}
	\min \max_{t \in {T}, j\in J}\{tx^t_{j, N}+p_j\}
\end{equation}
\text{subject to:}
{\small
\begin{eqnarray}
\text{(\ref{eqn:init_graph})--(\ref{eqn:band_in})}\nonumber\\
x^t_{j, n} \in \{0,1\}, \forall j\in J, n\in[N], t\in[T]\\
k_{(j,n)\rightarrow (j',n')}^t \geq 0, \forall j\in J, n\in[N], t\in[T], 
	 \nonumber\\ 
(j',n')\in succ(j,n), \text{$j$ and $j'$ are placed on different servers}
\end{eqnarray}
}
\vspace{-4mm}

}

Problem (\ref{eqn:flow_schedule_problem}) 
is a generalization of the strongly NP-hard multi-stage coflow scheduling problem (MSCSP)~\cite{tian2018scheduling}, by grouping transmission between the same types of tasks in one iteration as one coflow (\eg,~data transmission from all graph store servers to all samplers). In addition, the key challenge with our problem lies in the unknown graph data volume transferred between graph store servers and samplers: 
graph sampling is typically a random algorithm \cite{zeng2019graphsaint}, the training nodes and their neighbors selected vary from one training iteration to the next, 
and hence the sizes of node/edge features to transfer 
 change and are unknown beforehand. 
  Consequently, our execution scheduling is an online problem. 

In the following, we first design an online algorithm for task execution and flow transmission schedule, assuming task placements are given; next, we devise the task placement scheme that minimizes the total training time in conjunction with scheduling. An example task and flow schedule is given in Fig.~\ref{fig:gnn_planning}(b), where we depict task execution and flow communication for the first three training iterations, based on the task placement in Fig.~\ref{fig:gnn_planning}(a). Each training iteration is denoted using a different color. 

Key notation is summarized in Table~\ref{table:notation} for ease of reference.

\begin{table}
\caption{Notation}\label{table:notation}
\centering
\begin{tabular}{| c | l |}
\hline
$T$ & total time span\\
\hline
$J$ & set of all tasks\\
\hline
$J_{g}/J_{s}/J_{w}/J_{ps}$ & set of graph store servers/samplers/workers/PSs\\
\hline
$M$ & \# of machines\\
\hline
$N$ & \# of training iterations\\
\hline
$R$ & \# of resource types\\
\hline
$C^r_m$ & available amount of type-$r$ resource on machine $m$\\
\hline
$B^m_{in}~(B^m_{out})$ & avail. incoming (outgoing) bandwidth of machine $m$\\
\hline
$p_j$ & execution time of task $j$ in one iteration\\
\hline
$d_{(j,n)\rightarrow(j',n')}$ & amount of traffic 
transmitted from task $j$ of\\&   iteration $n$
to its successor task $j'$ of iteration $n'$ \\
\hline
$w^r_j$ & type-$r$ resource demand of task $j$\\
\hline
$y^m_j$ & task $j$ is placed on machine $m$ (1) or not (0)\\
\hline
$x_{j,n}^t$ & task $j$ of iteration $n$ starts at $t$ (1) or not (0)\\
\hline
$k_{(j,n)\rightarrow (j',n')}^t$ & amount of traffic transmitted from task $j$ of \\
& iteration $n$ to its successor $j'$ of iteration $n'$ at $t$ \\
\hline
\end{tabular}
\end{table}


%% file: sec/alg_flow.tex
\section{Online Execution and Flow Scheduling}
\label{sec::schedule}

\subsection{Scheduling Algorithm} 

Given placements $\{y_j^m\}$, we design an online algorithm that decides start time of each task ($x^t_{j,n}$) and flow transmission ($k_{(j,n)\rightarrow (j',n')}^t$) over time.

We maintain two flow sets: 
(i) $F_{act}$, that stores every active flow $(j,n)\rightarrow(j',n')$ which currently has started but not finished transmission yet; (ii) $F_{pend}$, to store every pending flow $(j,n)\rightarrow(j',n')$ whose predecessor task $(j,n)$ has been done, and that has not started because its predecessor flow $(j,n-1)\rightarrow(j',n'-1)$ in the previous iteration has not completed transmission yet.
For each task $(j,n)$, we use $\mathcal{F}(j,n)$ to represent the set of flows that originate from $(j,n)$ to tasks that reside on other machines (than where $j$ is).

\begin{algorithm}[!th]
\DontPrintSemicolon
\KwIn{$T, J, M, N, \{y_j^m\}$}
\KwOut{$\{x_{j,n}^t\}, \{k_{(j,n)\rightarrow (j',n')}^t\}, T_{OES}$}

Initialize $F_{act}$ and $F_{pend}$ to $\emptyset$\;
$x_{j,1}^1 \leftarrow 1, \forall j \in J_g$\;
\For{$t \in [T]$} 
{	
	\If {every $(j, N), j\in J$ is done (aka training has converged)}
	{
		$T_{OES} \leftarrow t-1$
		\Break
	}
	\For {$(j,n) \in \{(j,n)|j\in J,n\in[N]\}$} 
	{
		$x_{j,n}^t \leftarrow 1$ if $(j,n)$ is available\;
		\If {$(j,n)$ finished at $t-1$}
		{
			\For {$(j,n)\rightarrow (j',n') \in \mathcal{F}(j,n)$}
			{
				\If {\small $(j,n-1) \rightarrow (j',n'-1) \in F_{act} \cup F_{pend}$}
				{
					add $(j,n)\rightarrow (j',n')$ to $F_{pend}$\; 
				}
				\Else
				{
					add $(j,n)\rightarrow (j',n')$ to $F_{act}$\;
				}
			}
		}
	}
	\For {every flow $(j,n)\rightarrow (j',n')$ finished at $t-1$}
	{
		\If {$(j,n+1)\rightarrow (j',n'+1)\in F_{pend}$}
		{
			remove $(j,n+1)\rightarrow (j',n'+1)$ from $F_{pend}$\;
			add $(j,n+1)\rightarrow (j',n'+1)$ to $F_{act}$\;
		}
	}
	\For {$m \in [M]$}
	{
		calculate $\Delta_{in}^m$ and $\Delta_{out}^m$ according to (\ref{eqn:ingress_degree}) (\ref{eqn:egress_degree})\;
	}
	\For {$(j,n)\rightarrow (j',n') \in F_{act}$}
	{
		$k_{(j,n)\rightarrow (j',n')}^t \leftarrow \min\{B^{m'}_{in}/\Delta_{in}^{m'}, B^m_{out}/\Delta_{out}^m\}$, where $y^m_j = 1$ and $y^{m'}_{j'} = 1$
	}
}
\Return $\{x_{j,n}^t\}, \{k_{(j,n)\rightarrow (j',n')}^t\}, T_{OES}$\;
\caption{Online Execution Scheduling - \textbf{{OES}}}
\label{alg:execution_schedule}
\end{algorithm}
\setlength{\textfloatsep}{0pt}


Our online scheduling algorithm is in Alg.~\ref{alg:execution_schedule}. 
We start by running graph store server processing for the first training iteration at $t=1$ (line 2). Then at each time $t$, we run every task that has received all required data and hence is available to execute (line 7). For each task $(j,n)$ completed at $t-1$, consider every flow $(j,n)\rightarrow (j',n') \in \mathcal{F}(j,n)$ in $t$: if the flow's predecessor flow $(j,n-1)\rightarrow(j',n'-1)$ is in $F_{act}$ or $F_{pend}$ (indicating it not done yet), we add $(j,n)\rightarrow (j',n')$ to $F_{pend}$; otherwise, it is scheduled to transmit in $t$ and added to $F_{act}$ (lines 8-13). In addition, for every flow $(j,n)\rightarrow (j',n')$ ended at $t-1$, we check if its successor flow $(j,n+1)\rightarrow (j',n'+1)$ is in $F_{pend}$: if so, we move it from $F_{pend}$ to $F_{act}$ and start the flow transmission (lines 14-17).
For every flow $(j,n)\rightarrow (j',n')$ which transfers in $t$, supposing $j$ placed on $m$ and $j'$ on $m'$, we set its traffic volume $k_{(j,n)\rightarrow (j',n')}^t$ at $t$ to $\min\{B^{m'}_{in}/\Delta_{in}^{m'}, B^m_{out}/\Delta_{out}^m\}$ (lines 18-21). 
$\Delta_{in}^m$ ($\Delta_{out}^m$) is the \textit{ingress flow degree} (\textit{egress flow degree}) on machine $m$, 
counting the number of active flows entering and exiting from $m$, respectively: 

\vspace{-5mm}
{\small
\begin{equation}
	\label{eqn:ingress_degree}
	\Delta_{in}^m = |\{(j',n')\rightarrow (j,n)|(j',n')\rightarrow (j,n)\in F_{act}, y_{j}^m = 1\}|
\end{equation}
\begin{equation}
	\label{eqn:egress_degree}
	\Delta_{out}^m = |\{(j,n)\rightarrow (j',n')|(j,n)\rightarrow (j',n')\in F_{act}, y_j^m = 1\}|
\end{equation}}
\vspace{-5mm}

\noindent In this way, we balance flow rates among flows going into and out of each machine, ensuring no individual flow becoming the bottleneck. 
The algorithm terminates when the whole training process is done, \ie, all tasks of the last training iteration are completed 
(lines 4-5).



\subsection{Theoretical Analysis}

Let $F_{one\_iter}$ denote the set of all inter-machine flows in one training iteration, including the transfer of updated parameters computed in this iteration from PS to workers. We define the \textit{one-iteration ingress flow degree} $\widehat{\Delta^m_{in}}$ and \textit{one-iteration egress flow degree} $\widehat{\Delta^m_{out}}$: 

\vspace{-4mm}
{\small
\begin{eqnarray*}
	\widehat{\Delta^m_{in}} = |\{(j',n')\rightarrow (j,n)|(j',n')\rightarrow (j,n)\in F_{one\_iter}, y_j^m = 1\}|\label{eqn:one_ingress_degree}\\
	\widehat{\Delta^m_{out}} = |\{(j,n)\rightarrow (j',n')|(j,n)\rightarrow (j',n')\in F_{one\_iter}, y_j^m = 1\}|\label{eqn:one_egress_degree}
\end{eqnarray*}
}
\vspace{-5mm}

\noindent and the \textit{maximum degree} $\Delta$:
{\small
\begin{equation}
	\Delta = \max_{m\in[M]}\{\max\{\widehat{\Delta^m_{in}}, \widehat{\Delta^m_{out}}\}\}
	\label{eqn:delta}
\end{equation}}
\vspace{-5mm}

\noindent which represents the maximum number of incoming or outgoing flows at any machine in one training iteration. 

\begin{lemma}
	In any time step $t$, $\Delta_{in}^m$ ($\Delta_{out}^m$) are no larger than $\widehat{\Delta^m_{in}}$ ($\widehat{\Delta^m_{out}}$), for any $m \in [M]$.
	\label{lemma:schedule}
\end{lemma}

The detailed proof is given in Appendix~\ref{proof:lemma:schedule}.

\begin{theorem}
	The overall training makespan achieved by Alg.~\ref{alg:execution_schedule}, $T_{OES}$, is no larger than $\Delta$ times the 
	optimal objective value $T^*$ of the offline execution scheduling problem (\ref{eqn:flow_schedule_problem}), \ie, the competitive ratio of the online algorithm in Alg.~\ref{alg:execution_schedule} is $\Delta$.
	\label{theorem:schedule}
\end{theorem}

The detail proof is given in Appendix~\ref{proof:theorem:schedule}.

%% file: sec/alg_placement.tex
\section{Exploratory Task Placement}
\label{sec::placement}


We adopt the Markov Chain Monte Carlo (MCMC) search framework~\cite{geyer1992practical} to identify a good placement solution to minimize the training makespan with our online scheduling Alg.~\ref{alg:execution_schedule}. We start by constructing a feasible initial placement solution, $\mathcal{Y}_0 = \{y_j^m\}_0$, followed by generating a sequence of placements $\mathcal{Y}_1, \mathcal{Y}_2, \ldots$, until a time budget $I$ is exhausted.

\subsection{Constructing Initial Feasible Placement}

A feasible task placement solution should respect resource capacity constraints in~(\ref{eqn:device_capacity}). 
We first randomly order the $M$ machines 
into $\{m_1, m_2, \ldots, m_{M}\}$. Note that placements of graph store servers are given (one on a machine). 
Let 
$[q_s, q_w, q_{ps}, i]$ indicate that we can pack $q_s$ samplers, $q_w$ workers and $q_{ps}$ PSs within the first $i$ machine ($m_1$ to $m_i$) without violating resource capacities, and $(q_s, q_w, q_{ps}, i)$ be a particular partial placement of putting $q_s$ samplers, $q_w$ workers and $q_{ps}$ PSs on machine $m_i$. We use $\mathcal{A}_{q_s, q_w, q_{ps}, i}$ to denote an exact placement associated with $[q_s, q_w, q_{ps}, i]$, specifying how many samplers, workers and PSs are placed in each of the $i$ machines, to make up for the total numbers of $q_s$, $q_w$ and $q_{ps}$. Let $\Omega(i)$ be the set of all $[q_s, q_w, q_{ps}, i]$'s with $i$ fixed and $q_s\in [|J_s|], q_w\in [|J_w|], q_{ps}\in [|J_{ps}|]$. 

\begin{algorithm}
\DontPrintSemicolon
\KwIn{$J, M, R, \{C^r\}, \{w^r_j\}$}
\KwOut{$\mathcal{Y}_0$}
Randomly order $m$ machine as $\{m_1, m_2, \ldots, m_M\}$
$\Omega(m) \leftarrow \emptyset, \forall m\in[M]$\;
\For {$q_s,q_w,q_{ps} \in [\eta_s], [\eta_w], [\eta_ps]$}
{
	\If {$q_s$ samplers, $q_w$ workers, and $q_{ps}$ parameter servers can be packed in $m_1$}
	{
		add $(q_s, q_w, q_{ps}, 1)$ to set $\Omega(1)$\;
		$\mathcal{A}(q_s, q_w, q_{ps}, 1)\leftarrow\{<q_s, q_w, q_{ps}, 1>\}$\;
	}
}
\For {$i \in \{2, 3, \ldots, M\}$}
{	
	\For {$(q_s, q_w, q_{ps}, i-1) \in \Omega(i-1)$}
	{
		\If {$|J_s| - q_s$ samplers, $|J_w| - q_w$ workers, and $|J_{ps}| - q_{ps}$ parameter servers can be packed in $m_i$}
		{
			{\small $\mathcal{A}_{solution} \leftarrow \mathcal{A}(q_s, q_w, q_{ps}, i-1) \cup
			{\{<|J_s| - q_s, |J_w| - q_w, |J_{ps}| - q_{ps}, i>\}}$\;}
			Generate the initial placement solution $\mathcal{Y}_0 = \{y_j^m\}_0$ based on $\mathcal{A}_{solution}$\;
			\Return $\mathcal{Y}_0$
		}
		\For {$q'_s,q'_w,q'_{ps} \in [\eta_s], [\eta_w], [\eta_ps]$}
		{	
			\If {$q'_s$ samplers, $q'_w$ workers, and $q'_{ps}$ parameter servers can be packed in $m_i$}
			{
			\If {$q_s + q'_s \leq |J_s|$ and $q_w + q'_w \leq |J_w|$ and $q_{ps} + q'_{ps} \leq |J_{ps}|$}
			{
				add $(q_s + q'_s, q_w + q'_w, q_{ps} + q'_{ps}, i)$ to set $\Omega(i)$\;
				{\small$\mathcal{A}(q_s + q'_s, q_w + q'_w, q_{ps} + q'_{ps}, i)\leftarrow{\mathcal{A}(q_s, q_w, q_{ps}, i-1)} \cup {\{<q'_s, q'_w, q'_{ps}, i>\}}$}\;
			}
			}
		}
	}
}
\caption{Initial Placement Solution Construction - \textbf{\textit {IPSC}}}
\label{alg:init_solution}
\end{algorithm}

We use dynamic programming to construct a feasible placement solution. We first consider all feasible 
placements $(q_s, q_w, q_{ps}, 1)$ on $m_1$. 
Let $\eta_s$ denote the maximal number of samplers 
that can be hosted by any machine, \ie~$\eta_s = \max_{m\in[M]}\min_{r\in [R]:w^r_j > 0}\lfloor C'^r_m/w^r_j \rfloor,$ any $j\in J_{s}$ ($C'^r_m$ is available type-$r$ resource on $m$ excluding that occupied by the graph store server). Similarly, we can define an upper bound on the number of workers and PSs per machine, $\eta_w$ and $\eta_{ps}$. For every possible combination of $q_s\in \{0\}\cup[\min\{|J_s|, \eta_s\}]$, $q_w\in \{0\}\cup[\min\{|J_w|, \eta_w\}]$ and $q_{ps}\in \{0\}\cup[\min\{|J_{ps}|, \eta_{ps}\}]$, 
we check if the 
capacity constraints on $m_1$ are satisfied. For every feasible solution found, we add $[q_s, q_w, q_{ps}, 1]$ to $\Omega(1)$, and set 
$\mathcal{A}_{q_s, q_w, q_{ps}, 1}=\{(q_s, q_w, q_{ps}, 1)\}$. 

Next, we iteratively construct $\Omega(i)$ based on $\Omega(i-1)$ until finding a complete feasible solution of placing all $|J_s|$ samplers, $|J_w|$ workers and $|J_{ps}|$ PSs onto the machines. For each $[q_s, q_w, q_{ps}, i-1] \in \Omega(i-1)$, we examine whether $|J_s| - q_s$ samplers, $|J_w| - q_w$ workers and $|J_{ps}| - q_{ps}$ PSs can be fit into machine $m_i$. If so, we have identified a complete feasible placement solution that packs all tasks within the first $i$ machines: $\mathcal{A}_{solution} = \mathcal{A}(q_s, q_w, q_{ps}, i-1) \cup \{(|J_s| - q_s, |J_w| - q_w, |J_{ps}| - q_{ps}, i)\}$. 
Otherwise, we find every feasible placement $(q'_s, q'_w, q'_{ps}, i)$ with  $q'_s\in \{0\}\cup[|J_s| - q_s]$, $q'_w\in \{0\}\cup[|J_w| - q_w]$, and $q'_{ps}\in \{0\}\cup[|J_{ps}| - q_{ps}]$ that satisfies capacity constraint on machine $m_i$; and if $[q_s + q'_s, q_w + q'_w, q_{ps} + q'_{ps}, i]$ is not in $\Omega(i)$ yet, we add it into $\Omega(i)$, 
and set $A(q_s + q'_s, q_w + q'_w, q_{ps} + q'_{ps}, i)$ to be the union of $A(q_s, q_w, q_{ps}, i-1)$ and $\{(q'_s, q'_w, q'_{ps}, i)\}$. We build from $\Omega(2)$ to $\Omega(M)$ and return the first complete feasible placement solution. 
We summarize our initial placement solution construction algorithm in Alg.~\ref{alg:init_solution}, referred to as \textbf{IFS}. 

\vspace{-1mm}
\begin{theorem}
	\textbf{IFS} identifies a feasible placement solution within polynomial time.
	\label{theorem:init_placement}
\end{theorem}
\vspace{-2mm}

The detailed proof is given in Appendix~\ref{proof:theorem:init_placement}.

\begin{algorithm}[!t]
\DontPrintSemicolon
\KwIn{$T, J, M, R, \{C^r_m\}, \{w^r_j\}$}
\KwOut{$\mathcal{Y}_{min}$}

$\mathcal{Y}_0 \leftarrow $\textbf{{IFS}}($J, M, R, \{C^r_m\}, \{w^r_j\}$); $\mathcal{Y}_{min} \leftarrow \mathcal{Y}_0$\;
$\_, \_, min\_makespan = $~\textit{\bf OES}$(T, J, M, N, \mathcal{Y}_0)$\;
\For {$z \in \{0, 1, \ldots, I-1\}$}
{
	randomly select a task $j$ from $J\setminus J_g$\;
	construct $M_{avail}$ of $j$\;
	randomly select a machine $m$ from $M_{avail}$\;
	construct new placement solution $\mathcal{Y}'$\;
	$\_, \_, T'_{\mathcal{Y}'} = $~\textit{\bf OES}$(T, J, M, N, \mathcal{Y}')$\;
	$\pi(\mathcal{Y}_z \rightarrow \mathcal{Y}')\leftarrow \min\{1, \exp(\beta cost(\mathcal{Y}_z) - \beta cost(\mathcal{Y}'))\}$\;
	\uIf{rand()$ \leq \pi(\mathcal{Y}_z \rightarrow \mathcal{Y}')$}
	{
		$\mathcal{Y}_{z+1} \leftarrow \mathcal{Y}'$\;
		\If {no resource violation with $\mathcal{Y}_{z+1}$ and $T'_{\mathcal{Y}'} < min\_makespan$}
		{	
			$min\_makespan \leftarrow T'_{\mathcal{Y}'}$\;
			$\mathcal{Y}_{min} \leftarrow \mathcal{Y}_{z+1}$\;
		}
	}
	\Else
	{	
		$\mathcal{Y}_{z+1} \leftarrow \mathcal{Y}_z$\;
	}
}
\Return $\mathcal{Y}_{min}$
\caption{Exploratory Task Placement - \textbf{{ETP}}}
\label{alg:mcmc}
\end{algorithm}
\setlength{\textfloatsep}{0pt}

\vspace{-1mm}
\subsection{Searching for Better Placements}
\label{subsec:cost}
\vspace{-1mm}

Starting from the initial feasible placement, we 
iteratively search for better placement solutions, 
according to a $cost(\mathcal{Y})$ 
defined on the overall training makespan of placement $\mathcal{Y}$.

Practically, task placements should be decided before training starts and remain fixed during training (to avoid substantial overhead of VM/container migration and flow redirection). The online nature of execution scheduling is due to size variation of sampled graph data; we should identify a placement that works best in expectation of the traffic variation.
To this end, we profile task execution time and inter-task traffic volumes by running the GNN training for some iterations ($50$ as in our evaluation), and produce their distributions. We simulate the training process under placement $\mathcal{Y}$ 
using Alg.~\ref{alg:execution_schedule}, with time and traffic volume drawn from the distributions, and derive the training makespan $T'_{\mathcal{Y}}$. 
The cost of placement $\mathcal{Y}$ 
is: 

\vspace{-3mm}
{\small
\begin{equation}
	cost(\mathcal{Y}) = T'_{\mathcal{Y}}(1 + \sum\limits_{m\in[M], r\in[R]} \max\{\frac{\sum\limits_{j\in J}w^r_jy^m_j - C^r_m}{C^r_m}, 0\})
	\label{eqn:cost}
\end{equation}
}

\noindent where $T'_{\mathcal{Y}}$ is multiplied by 1 plus a penalty term for resource violation (computed as the sum of capacity violation percentages over all types of resources and all machines).



Our search explores the solution space by transferring from one placement $\mathcal{Y}_z$ to another $\mathcal{Y}_{z+1}$, for a total of $I$ transfers (the time budget). 
Give $\mathcal{Y}_z$, 
we 
uniformly randomly sample a task $j \in J\setminus J_g$.
Let $M_{avail}$ denote the set of machines other than the one where $j$ is placed in $\mathcal{Y}_z$, which can host $j$ adhering to relaxed resource capacity constraints: 

\vspace{-2mm}
{\small
\begin{equation}
	\sum\limits_{j\in J}w^r_jy^m_j \leq (1+\mu)C^r_m, \forall m\in [M], r\in[R]
	\label{eqn:relaxed_capacity}
\end{equation}}
\vspace{-3mm}

\noindent Here, the capacity constraints are relaxed by a $\mu$ factor to allow full exploration in the placement space. 
For example, when the violation factor $\mu$ is set to $100\%$ (default in our evaluation), 
every feasible solution can be identified if an infinite time budget $I$ is allowed: 
Setting $\mu$ to $100\%$ is equivalent to allowing a duplicate set of machines (\ie,~each machine has doubled its resource capacities).
Therefore, we can transit from any feasible placement $\mathcal{Y}$ to any other feasible $\hat{\mathcal{Y}}$ by moving each task from its placement in $\mathcal{Y}$ to the duplicate of the machine where it is placed in $\hat{\mathcal{Y}}$. The new placement on the set of duplicate machines is feasible since 
$\hat{\mathcal{Y}}$ is a feasible placement solution. 
If the computational resources of all machines are quite sufficient to host the training tasks, we can set $\mu$ to a smaller value for better search efficiency.

 Next, we uniformly randomly choose one server $m \in M_{avail}$ to move $j$ to, and come up with the new placement solution 
$\mathcal{Y}'$.
 We compute a probability ($\beta>0$ is a hyperparameter set to 0.1 in our evaluation, whose smaller value  increases the tendency of our search process to jump out of local optima): 

\vspace{-4mm}
{\small
\begin{eqnarray}
	\label{eqn:criteria}
	\pi(\mathcal{Y}_z \rightarrow \mathcal{Y}') 
	= \min\{1, \exp(\beta cost(\mathcal{Y}_z) - \beta cost(\mathcal{Y}'))\}
\end{eqnarray}}
\vspace{-4mm}

\noindent With probability $\pi(\mathcal{Y}_z \rightarrow \mathcal{Y}')$, we use $\mathcal{Y}'$ as $\mathcal{Y}_{z+1}$: if $cost(\mathcal{Y}') \leq cost(\mathcal{Y}_z)$, we accept $\mathcal{Y}'$ as 
$\mathcal{Y}_{z+1}$ (probability is 1); otherwise, we still accept $\mathcal{Y}'$ as the next state with probability $\pi(\mathcal{Y}_z \rightarrow \mathcal{Y}')$ (for exploration) and maintain $\mathcal{Y}_{z+1}$ the same as $\mathcal{Y}_z$ with probability $1 - \pi(\mathcal{Y}_z \rightarrow \mathcal{Y}')$. 

Our state transition as designed above 
ensures that the probability of visiting $\mathcal{Y}$ is linear to $\exp(-\beta cost(\mathcal{Y}))$~\cite{geyer1992practical},
\ie, solutions with lower costs are more frequently visited than ones with larger costs. We return the best feasible placement 
found after $I$ transitions, 
which does not violate any original resource capacity constraints in~(\ref{eqn:device_capacity}), and leads to the minimum (simulated) training time as compared to all other feasible placements visited.
Alg.~\ref{alg:mcmc} summarizes our exploratory task placement algorithm (\textbf{ETP}).




%% file: sec/alg_complete.tex
\subsection{Complete Distributed GNN Training Planning Algorithm}
\label{sec::complete}

\begin{algorithm}[!t]
\DontPrintSemicolon
\KwIn{$T, N,  J, M, R, \{C^r_m\}, \{w^r_j\}$}
\KwOut{$\mathcal{Y}_{min}, \{x_{j,n}^t\}, \{k_{(j,n)\rightarrow (j',n')}^t\}$}

$\mathcal{Y}_{min} \leftarrow$ \textbf{{ETP}}($T, J, M, R, \{C^r_m\}, \{w^r_j\}$)\;
$\{x_{j,n}^t\}, \{k_{(j,n)\rightarrow (j',n')}^t\}, T_{OES} \leftarrow$  \textbf{{OES}}($T, J, M, N, \mathcal{Y}_{min}$)\;
\Return $\mathcal{Y}_{min}, \{x_{j,n}^t\}, \{k_{(j,n)\rightarrow (j',n')}^t\}$\;

\caption{Distributed GNN Training Planning (\textit{{DGTP}})}
\label{alg:dgtp}
\end{algorithm}
\setlength{\textfloatsep}{0pt}

Our complete distributed GNN training planning (\textit{DGTP}) algorithm is given in Alg.~\ref{alg:dgtp}. We first leverage Alg.~\ref{alg:mcmc} to identify the best placement $\mathcal{Y}_{min}$ and then use Alg.~\ref{alg:execution_schedule} to decide the task and flow schedules $\{x_{j,n}^t\}, \{k_{(j,n)\rightarrow (j',n')}^t\}$ based on $\mathcal{Y}_{min}$, in an online manner. 


%% file: sec/eval.tex
\vspace{-2mm}
\section{Performance Evaluation}
\label{sec::eval}

We evaluate {\em DGTP} by both testbed experiments and simulation studies.

\subsection{Testbed Experiments}
\noindent\textbf{Implementation.} We implement {\em DGTP} using Python on DGL 0.6.1~\cite{wang2019deep} and PyTorch 1.8.1~\cite{adam2019pytorch} with 1056 LoC for the training system and 1522 LoC for the search and scheduling algorithms. 
Parameter synchronization through a PS is built on PyTorch. We use the Stochastic Fairness Queueing provided by tc qdisc~\cite{brown2006traffic} 
to control flow transmission rates according to our online scheduling algorithm, \revise{dynamically assigning ongoing data flows into separate queues and ensuring fairness among them with negligible scheduling overhead.}

\noindent\textbf{Testbed.} 
Our testbed consists of 4 GPU servers inter-connected by a Dell Z9100-ON switch, with 50Gbps peak bandwidth between any two servers. Each server is equipped with one 50GbE NIC, one 8-core Intel E5-1660 CPU, two GTX 1080Ti GPUs and 48GB DDR4 RAM. To emulate resource heterogeneity, we use tc to limit the bandwidth capacity of two servers to 10Gbps.

\noindent\textbf{GNN model and datasets.} We train one representative GNN model (three layers of hidden size 256), GraphSage~\cite{hamilton2017inductive}, on two graph datasets: ogbn-products~\cite{hu2020open} (an Amazon product co-purchasing graph) and Reddit~\cite{hamilton2017inductive} (consisting of Reddit posts within a month and connections between posts if the same user comments on both posts).
We implement uniformly random sampling of neighbors of training nodes, with different fan-outs (the number of neighbors to sample) at different hops, set 
according to the official training script 
provided by the DGL team and other existing studies \cite{hamilton2017inductive}\cite{zheng2020distdgl}. 
Same as in {\em DistDGL}~\cite{zheng2020distdgl}, we set the mini-batch size on both datasets to 2000 (subgraphs). We use Adam optimizer~\cite{goodfellow2016deep} with a learning rate of 0.001 during the training.

{\small
\begin{table}[!ht]
\caption{Benchmark datasets}
\label{tab:dataset}
\begin{tabular}{|c|c|c|c|c|}
\hline
Dataset                               & \#Nodes                          & \#Edges                            & \begin{tabular}[c]{@{}c@{}}Feature \\ Vector Length\end{tabular} & \multicolumn{1}{c|}{Fan-out}    \\ \hline
ogbn-products                         & 2.4M                        & 61.8M                         & 100                                                              & 5, 10, 15                       \\ \hline
Reddit                                & 0.2M                          & 114.6M                        & 602                                                              & 5, 10, 25                       \\ \hline
\multicolumn{1}{|c|}{ogbn-papers100M} & \multicolumn{1}{c|}{0.1B} & \multicolumn{1}{c|}{1.6B} & 128                                                              & \multicolumn{1}{c|}{12, 12, 12} \\ \hline
\end{tabular}
\end{table}}

We use 4 graph store servers, 6 workers (each requiring 3GB memory, 1 logical CPU core and 1 GPU card), and 1 PS (requiring 5GB memory, 1 logical CPU core)
to train the GNN model. We partition the input graph with METIS partitioner~\cite{karypis1998fast} among graph stores.
Each worker is associated with two samplers (each requiring 7GB memory, 2 logical CPU cores). 
We profile data to drive our search algorithm over $50$ iterations of training on each dataset. 

\noindent\textbf{Baseline.}
We compare {\em DGTP} with a modified version of {\em DistDGL}~\cite{zheng2020distdgl} that enables inter-server communication between a worker and its samplers. {\em DistDGL} adopts a placement scheme 
that maximally colocates each worker with its associated samplers within one server, 
and uses the system default scheduling strategy (running a task when ready and sending data in FIFO queues). 

\begin{figure*}[!t]
	\begin{minipage}[t]{0.50\textwidth}
    \begin{subfigure}{0.48\columnwidth}
      \centerline{\includegraphics[width=\columnwidth]{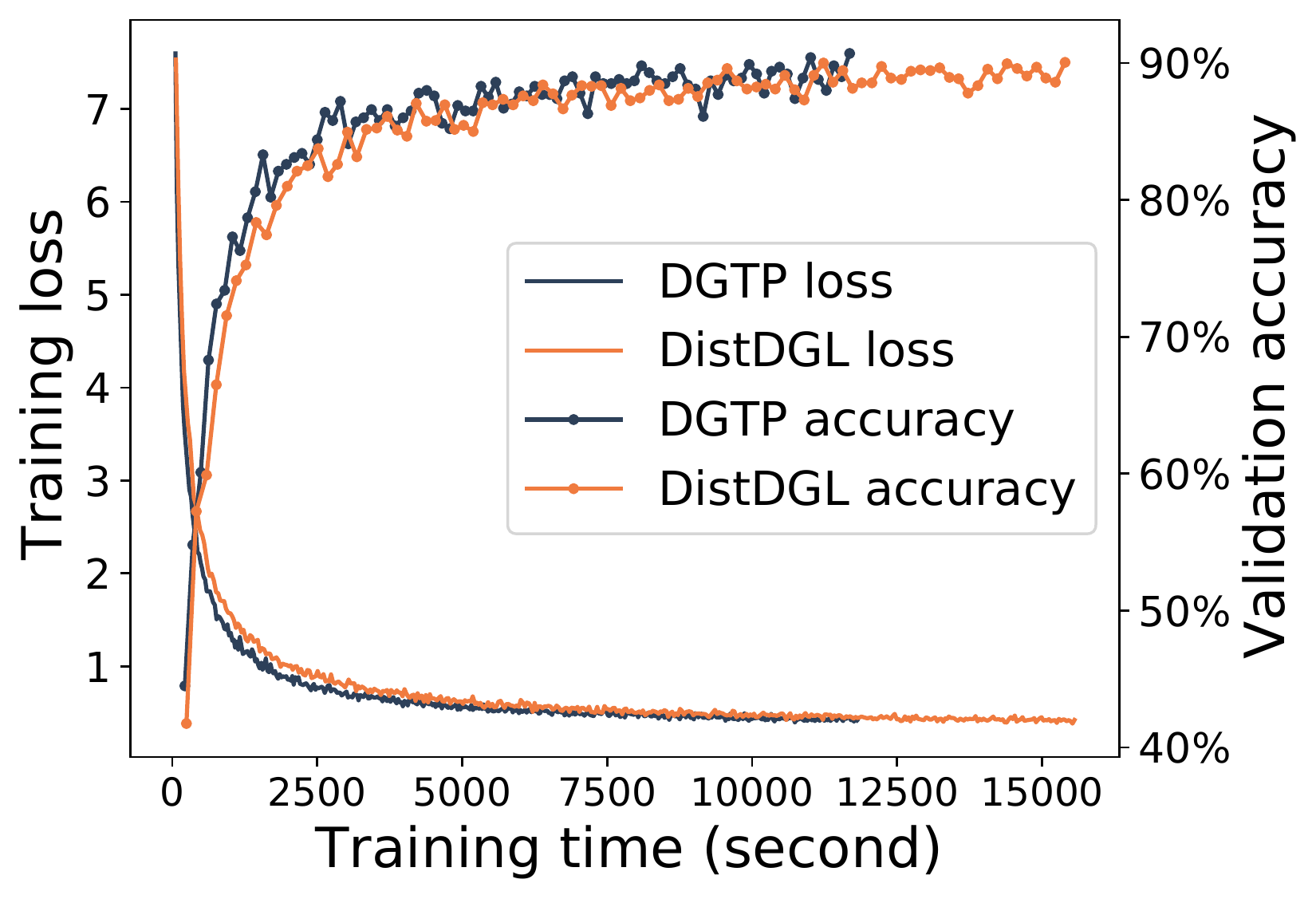}}
      \caption{ogbn-products}
    \end{subfigure}
    \begin{subfigure}{0.48\columnwidth}
      \centerline{\includegraphics[width=\columnwidth]{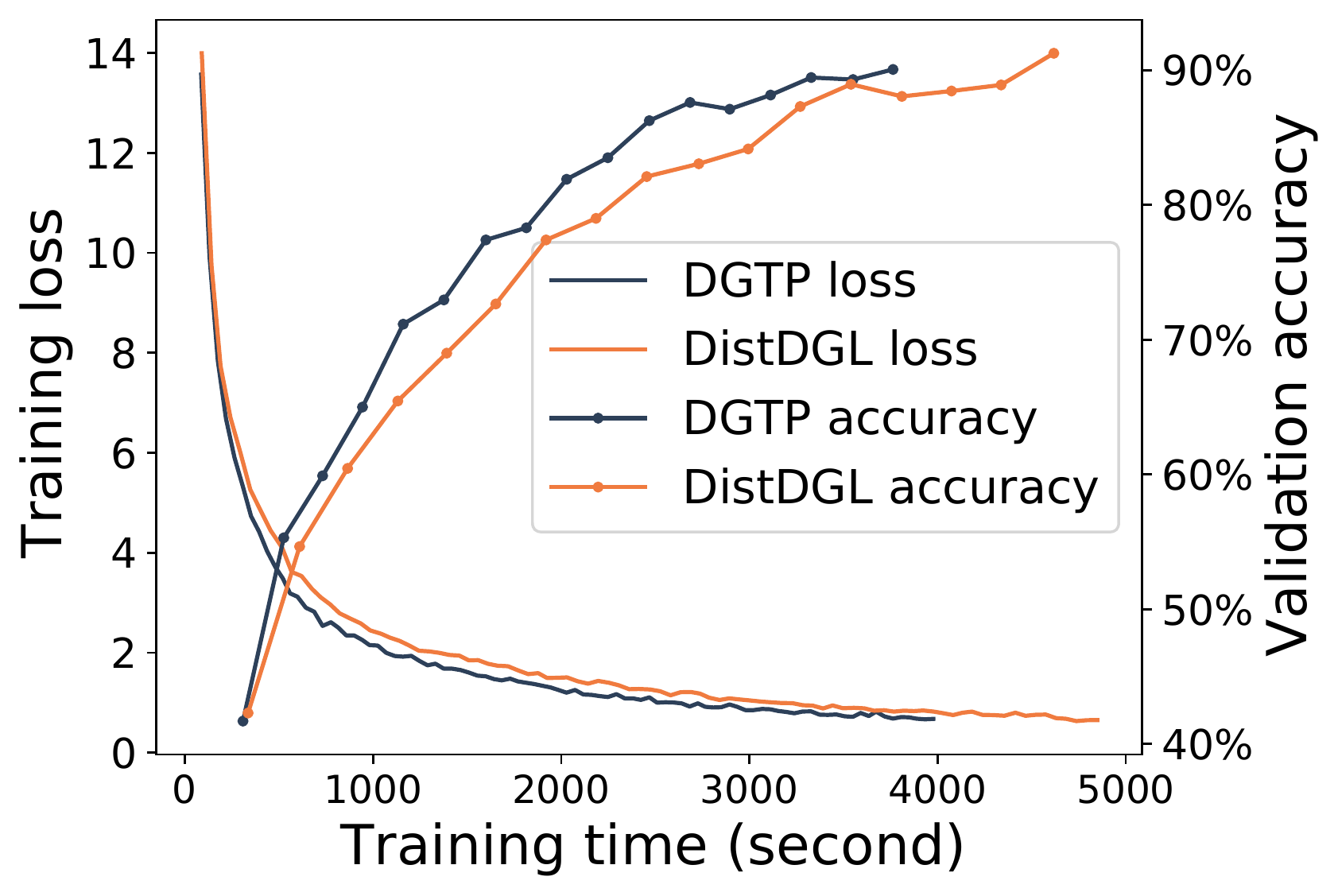}}
      \caption{Reddit}
    \end{subfigure}
    \caption{Training loss \& validation accuracy: {\em DGTP} vs. {\em DistDGL}}
    \label{fig_loss_acc}
    \end{minipage}
    \begin{minipage}[t]{0.50\textwidth}
    \begin{subfigure}{0.47\columnwidth}
      \centerline{\includegraphics[width=\columnwidth]{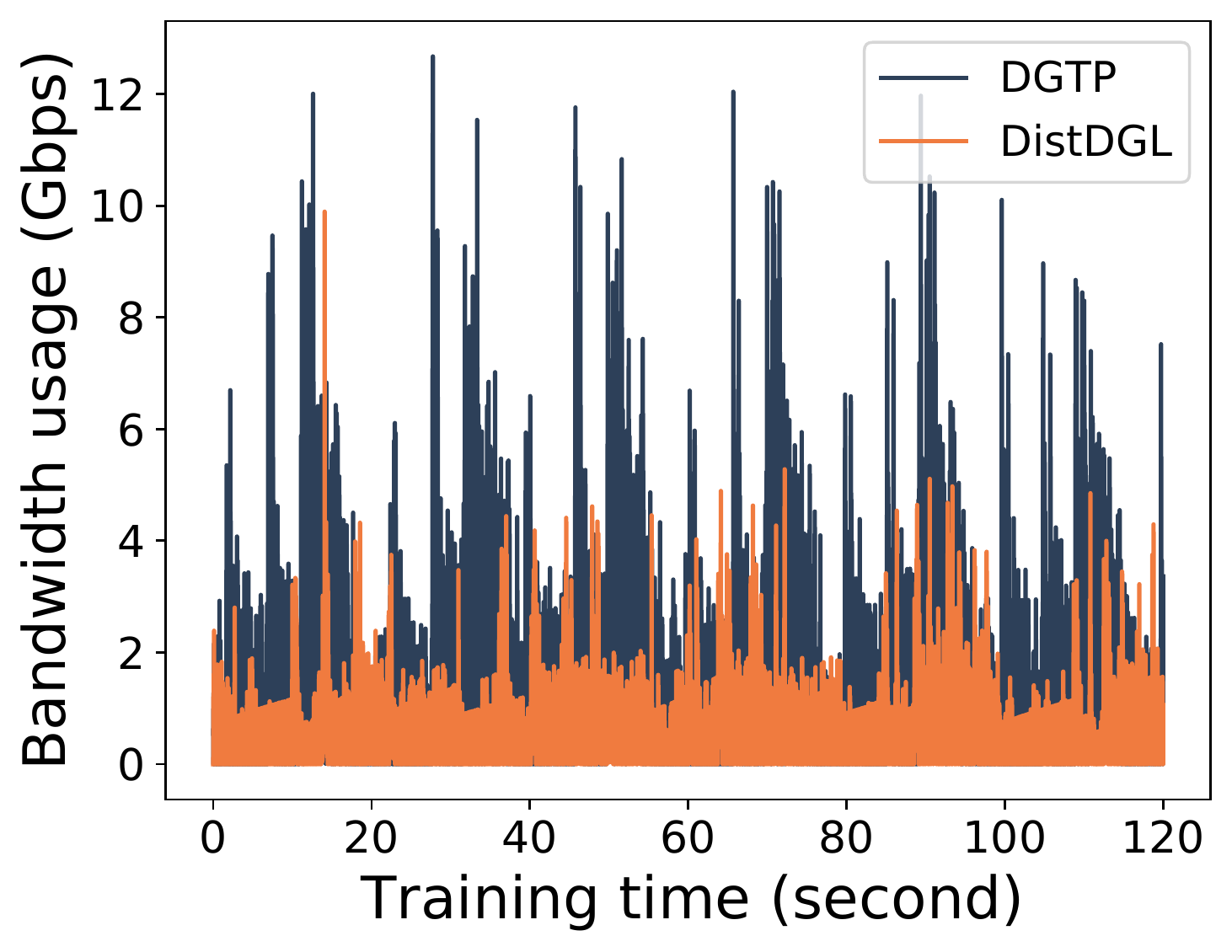}}
      \caption{ogbn-products}
    \end{subfigure}
    \begin{subfigure}{0.47\columnwidth}
      \centerline{\includegraphics[width=\columnwidth]{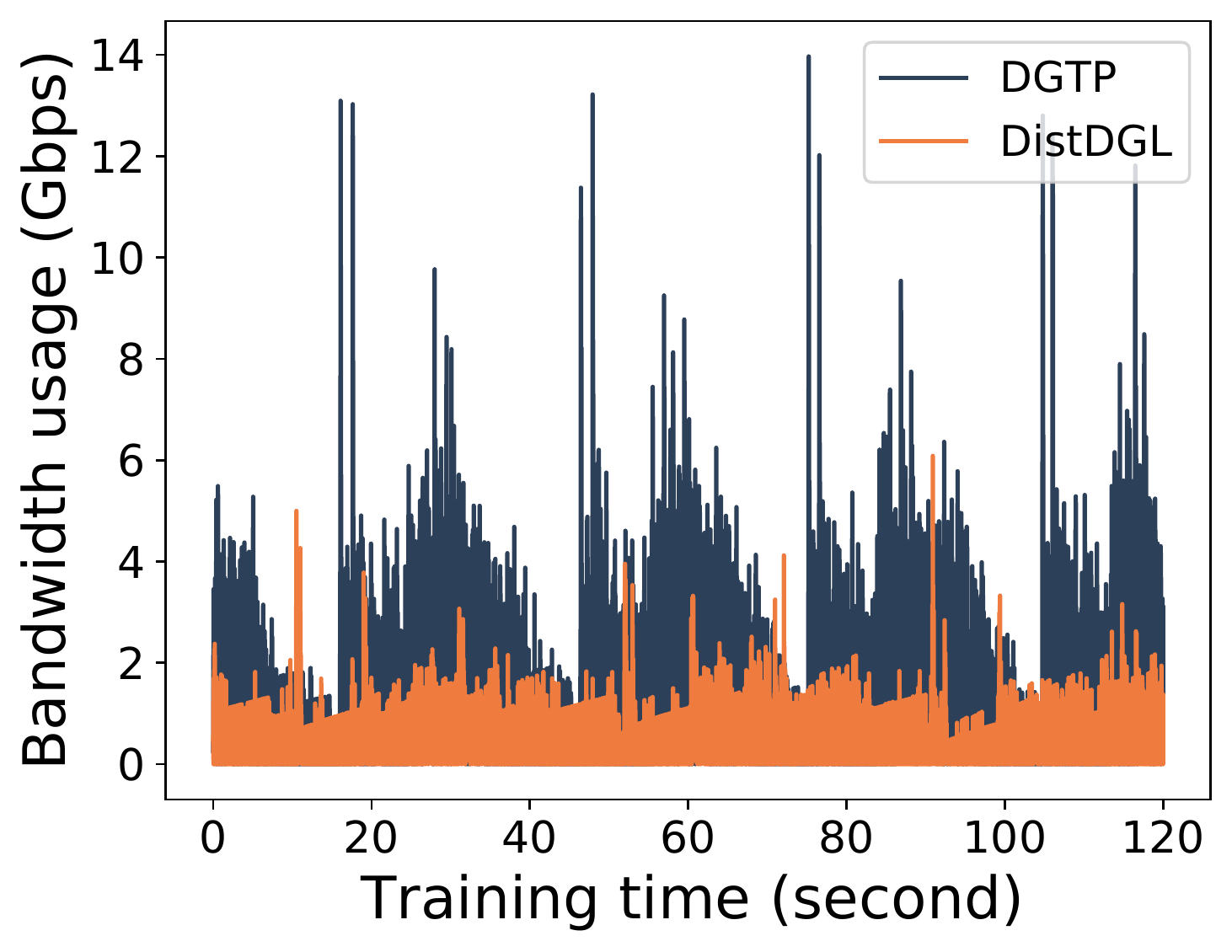}}
      \caption{Reddit}
    \end{subfigure}
    \caption{Total network bandwidth usage: {\em DGTP} vs. {\em DistDGL}}
    \label{fig_network_usage}
    \end{minipage}
    \begin{minipage}[t]{0.24\textwidth}
		\includegraphics[width=\textwidth]{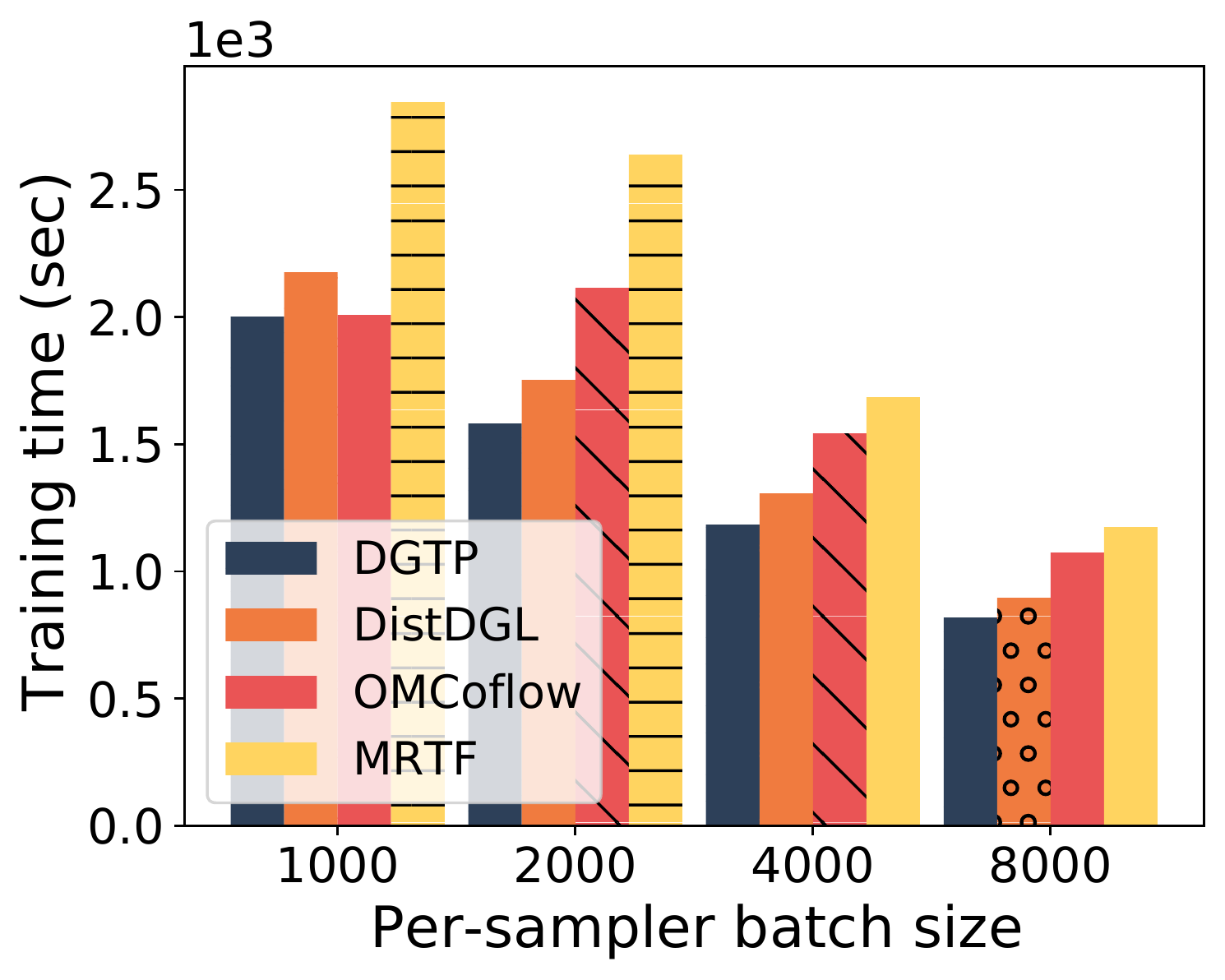}
		\caption{Training time on ogbn-products: different batch sizes}
		\label{diff_batch_products}
	\end{minipage}
	\begin{minipage}[t]{0.24\textwidth}
		\includegraphics[width=\textwidth]{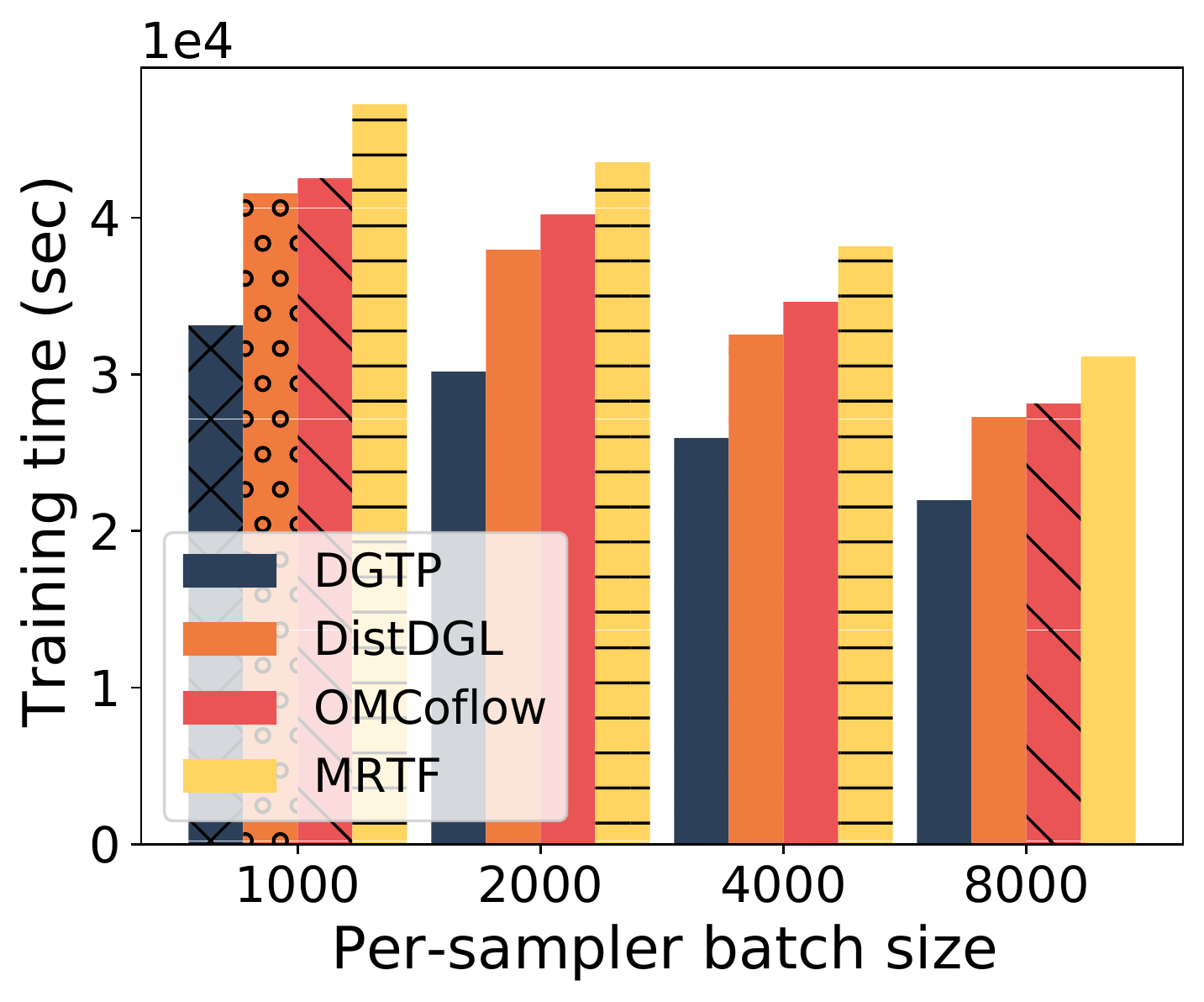}
		\caption{Training time on ogbn-papers100M: diff. batch sizes}
		\label{diff_batch_papers100M}
	\end{minipage}
	\begin{minipage}[t]{0.24\textwidth}
		\includegraphics[width=\textwidth]{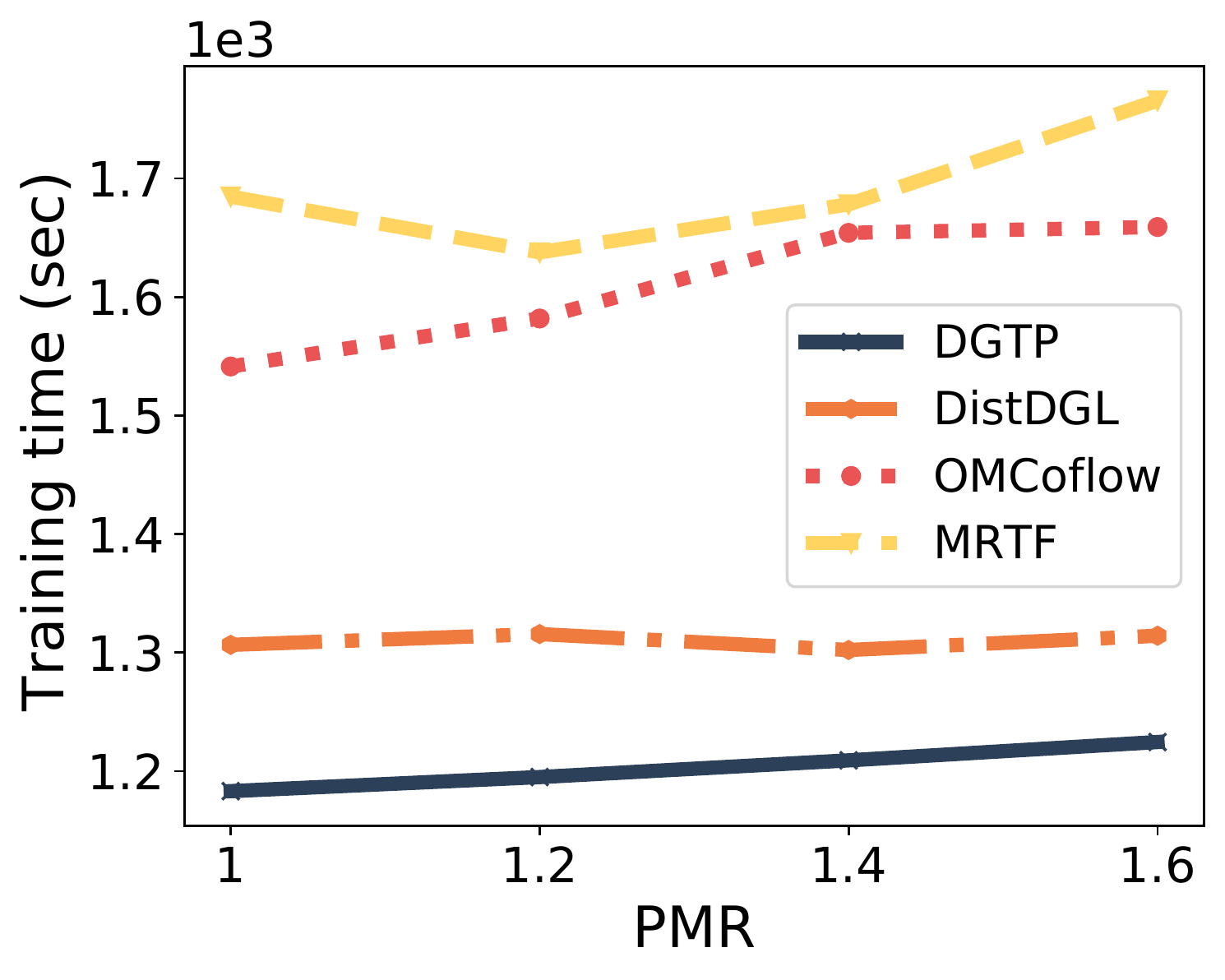}
		\caption{Training time on ogbn-products: different PMRs}
		\label{diff_pmr_products}
	\end{minipage}
	\begin{minipage}[t]{0.24\textwidth}
		\includegraphics[width=\textwidth]{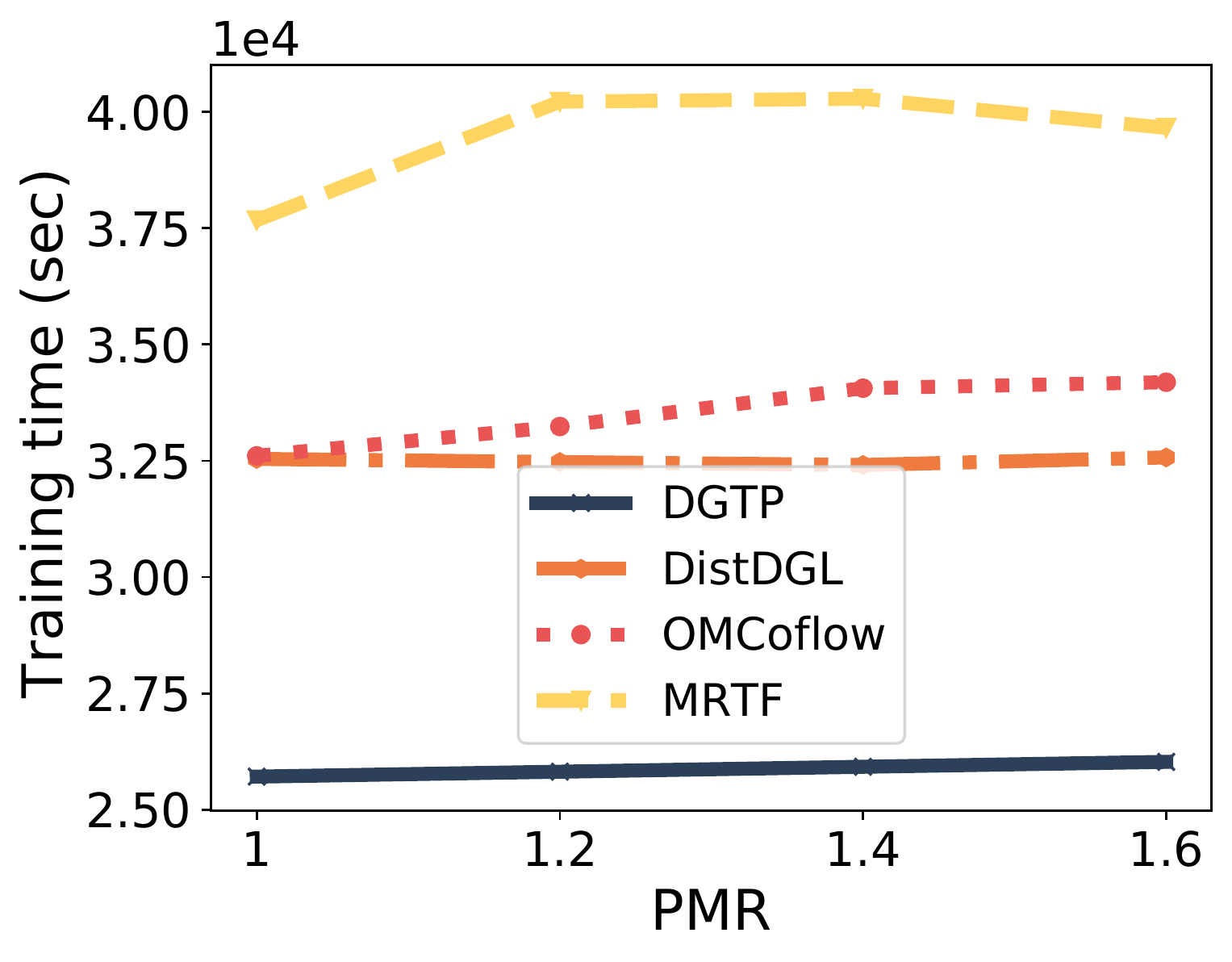}
		\caption{Training time on ogbn-papers100M: different PMRs}
		\label{diff_pmr_papers100M}
	\end{minipage}
	\vspace{-6mm}
\end{figure*}

\noindent\textbf{End-to-end training performance.} We compare the end-to-end training convergence time between {\em DGTP} and {\em DistDGL}. 
The offline search to obtain {\em DGTP}'s task placements can be done within 5 minutes (\revise{we only simulate 20 iterations of GNN training to obtain $T'_{\mathcal{Y}}$ during search, and set the exploratory time budegt to 10000}).
Fig.~\ref{fig_loss_acc} shows the training progress 
to achieve a 90\% target accuracy over the validation sets. {\em DGTP} outperforms {\em DistDGL} by 31.75\% in terms of the overall training time on ogbn-products, and 22.95\% on Reddit. 

\noindent\textbf{
Resource usage.} We also examined resource usage during training. We observed similar GPU, CPU and memory usage between {\em DGTP} and {\em DistDGL}, as task execution in both systems is constantly blocked by the large data transfers. Fig.~\ref{fig_network_usage} shows the 
bandwidth usage on the four servers. 
We observe that 
{\em DGTP} has a much better network usage on both datasets: 
{\em DGTP} can identify task placements that 
exploit the heterogeneous bandwidth levels well, 
while communication in {\em DistDGL} is often bottlenecked on the low-bandwidth inter-server connections 
(two pairs of its worker and samplers have to be separated onto different servers due to non-sufficient resources on the same servers).

\vspace{-1mm}
\subsection{Simulation Studies}
\vspace{-1mm}

\noindent\textbf{Settings.} We further evaluate {\em DGTP} in detail 
by simulating the training of the GraphSage model on: 1) ogbn-products on 8 machines using 8 graph store servers, 16 workers each with 2 samplers, and 1 PS; and 2) ogbn-papers100M (Microsoft Academic Graph dataset described in Table~\ref{tab:dataset}) on 16 machines using 16 graph store servers, 20 workers each with 4 samplers, and 1 PS. 
We simulate 5 epochs of training (\ie, each sampler goes through the whole set of training nodes specified by the dataset for five times) 
on ogbn-products (actual training of GraphSage on ogbn-products converges in 5 epochs, as we observed in our experiment), 
and 25 epochs on ogbn-papers100M (convergence time according to ogbn-papers100M official leaderboard~\cite{ogbnpapers100msage}). Our simulation is driven by profiled data collected by training the model on the respective datasets in our testbed. 


We consider four types of resources on each machine: memory, CPU, GPU and network bandwidth. The available memory size on each machine is set within $[32, 128]$GB, the number of available CPU cores between $[4, 16]$, the number of available GPUs within $[1, 4]$, and network bandwidth among $\{10Gbps, 20Gbps, 50Gbps\}$.

\noindent\textbf{Baselines.}
Apart from {\em DistDGL}, 
we further compare {\em DGTP} with two flow scheduling schemes (in which we use the same placements as computed by {\em DGTP} and a task starts immediately once its dependencies have been cleared):
(i) {\em OMCoflow}, a state-of-the-art online coflow scheduling algorithm~\cite{tan2019joint} that groups flows to the same task as one coflow, and sets the flow rates in each coflow inversely proportional to predicted flow finish time (supposing it is the only coflow in the network);
(ii) {\em MRTF}, which schedules flows according to the minimum remaining time first (MRTF) heuristic.


\noindent\textbf{Different per-sampler batch sizes.} 
A larger per-sampler batch size (a worker's mini-batch size divided by the number of samplers it uses) results in larger sampling data traffic, potentially yielding more communication overhead when poorly planned. As Fig.~\ref{diff_batch_products} and Fig.~\ref{diff_batch_papers100M} show, {\em DGTP} outperforms all three baselines, reducing the training makespan on ogbn-products by up to 11\%, and on ogbn-papers100M by up to 25\%, compared to {\em DistDGL}.
Larger data traffic is incurred for training on ogbn-papers100M due to the larger fan-outs, and its training environment is more complex (with more servers, resource heterogeneity, etc.). 
We identify {\em DGTP}'s larger speed-up on ogbn-papers100M is because {\em DGTP} can find better task placements that reduce the overall data traffic during training and schedule the traffic over the complex network environment well.
Further, {\em DGTP} achieves up to 33\% less training time as compared to {\em OMCoflow}, and up to 67\% to {\em MRTF}, on the two datasets. The advantage of {\em DGTP} 
improves with batch size. 
These indicate that {\em DGTP} can efficiently schedule 
flow transfers to minimize the overall training time in an online manner. 

\vspace{-1mm}
\noindent\textbf{Different peak-to-mean ratios.} 
We compute a peak-to-mean ratio (PMR) for flows from graph store servers to samplers, as the maximum data flow rate between any (graph store server, sampler) pair divided by the average flow rates among all such flows. 
The PMR in our profiled data during training with {\em DGTP} is 1.16 on ogbn-products and 1.08 on ogbn-papers100M. 
We scale up and down the transmitted graph data sizes to simulate different PMRs. Intuitively, a larger PMR indicates more intensive traffic volume fluctuation, more challenging for online scheduling. In Fig.~\ref{diff_pmr_products} and Fig.~\ref{diff_pmr_papers100M}, {\em DGTP} exhibits stable performance as the PMR changes, and outperforms {\em DistDGL} by up to 26\%, {\em OMCoflow} by up to 37\% and {\em MRTF} by up to 55\%.

%% file: sec/conclusion.tex
\vspace{-1mm}
\section{Conclusion}
\label{sec::conclusion}
\vspace{-1mm}

This paper designs efficient placement and scheduling algorithms for distributed GNN training over heterogeneous clusters. We propose a competitive online execution algorithm that schedules training task execution and flow transfers for both graph data sampling and parameter synchronization. We also design an explorative algorithm to decide the placement of every task, which, in conjunction with online task/flow scheduling, minimizes the overall training makespan. According to testbed experiments, our design reduces the end-to-end training time by up to 31.75\% as compared to a state-of-the-art distributed GNN training solution. Simulation studies further demonstrate that our design significantly outperforms representative schemes by minimizing the total data traffic and maximizing the bandwidth usage through task placement, and strategically scheduling tasks and flows to overlap computation with communication and reduce total communication time. 
Our design can be easily extended to GNN training with AllReduce-based parameter synchronization, by considering detailed communication flows within AllReduce operations. In the case of multiple GNN training jobs on the same cluster, our algorithms can be adopted for jointly searching for task placements of all jobs, and for online task and flow scheduling for each of the jobs. 

%% file: sec/appendix.tex
\appendices
\section{Proof of Lemma~\ref{lemma:schedule}}
\label{proof:lemma:schedule}
\begin{proof}
	Without loss of generality, let us consider a time step $t$. Due to the inter-iteration dependency (\ref{eqn:flow_component}), there will be at most one data flow from task $j$ to task $j'$. As a result, for any active flow $(j,n)\rightarrow (j',n')$ in $F_{act}$, we can substitute the flow with its counterpart one (\ie~the same flow from task $j$ to task $j'$) in the first iteration. Let us substitute all flows in $F_{act}$ with their counterparts in the first iteration and denote the new set as $F_{act}'$. Clearly, we can still obtain the same $\Delta_{in}^m$ and $\Delta_{out}^m$ with $F_{act}'$.
	
	Noting that $F_{one\_iter}$ includes all inter-server flows from the first iteration, we have $F_{act}' \subseteq F_{one\_iter}$. Consequently, $\Delta_{in}^m \leq \widehat{\Delta^m_{in}}$ and $\Delta_{out}^m \leq \widehat{\Delta^m_{out}}$.
\end{proof}

\section{Proof of Theorem~\ref{theorem:schedule}}
\label{proof:theorem:schedule}
\begin{proof}
	We aim at construct a chain $O: o_L \rightarrow o_{L-1} \rightarrow \dots \rightarrow o_1$, starting from the execution of one of the graph store servers in iteration 1 to the last parameter server in iteration $N$, strictly following either the execution dependency or the inter-iteration dependency. Every part in chain $O$ represents either a task or a data flow between tasks. We construct the chain reversely by setting $o_1$ as the last parameter server, denoted as $(j_{last}, N)$, in iteration $N$ that finishes at $T_{OES}$. We use $\tilde{t}(o_l)$ to denote the start time of $o_l, \forall l \in [L]$. Now, considering $\tilde{t}(o_1)$, there are two possibilities:
	
	\textbf{1):} All flows to $o_1$ finish before $\tilde{t}(o_1)$. 

	\textbf{2):} At least one flow to $o_1$ finishes at $\tilde{t}(o_1)$.
	
	\textbf{The first possibility} indicates that the execution of the same task for previous batch ($(j_{last}, N-1)$ in this case) finishes at $\tilde{t}(o_1)$. Therefore, we can add the execution of $(j_{last}, N-1)$ to chain $O$ denoted as $o_2$. Now we extend $O$ further by one task to cover more part of the total makespan $T_{OES}$.
	
	For \textbf{the second possibility}, without loss of generality, we consider one flow, denoted as $f_1$, to $o_1$ finishing at $\tilde{t}(o_1)$. We can add the flow $f_1$ to chain $O$ as $o_2$. Now, considering the start time of $f_1$, $\tilde{t}(f_1)$, there are two cases:
	
	$\triangleright$ \textbf{Case 1}: The task that $f_1$ originates from finished at $\tilde{t}(f_1)$.
	
	In this case, we can add the task to $O$ as $o_3$. We use $o_3$ and $o_2$ to further extend $O$, covering more makespan.
	
	$\triangleright$ \textbf{Case 2}: The task that $f_1$ originates from finished before $\tilde{t}(f_1)$.
	
	\textbf{Case 2} indicates that there exists another flow $f_2$ that blocks the transmission of $f_1$ due to the inter-iteration dependency. Now, we add both $f_1$ and $f_2$ to chain $O$ as $o_2$ and $o_3$. Consider $f_2$, there are again two cases as $f_1$. We can follow the same procedure to add either another flow $f_3$ or a task to $O$. We repeat the procedure until we add a task to $O$.
	
	In conclusion, we add one task and possibly several flows to extend our chain $O$ that the execution of one part starts immediately after the completion of its previous one. 
	For the newly added task, there are again two possibilities as $o_1$ and we can follow the same process as above to further extend the coverage of makespan by adding another task to $O$. Eventually, we can construct a chain $O$ to cover the entire makepan $T_{OES}$. 	
	Denote the total execution time of the task in $O$ as $p_{sum}$. 
	Supposing that there are $Q$ data transmissions in $O$, we use $d_1$ to $d_Q$ to denote the amount of data for each transmission. In addition, we use $m^{in}_q$ ($m^{out}_q$) to denote the server where the $q$-th flow comes to (from). 
	Clearly, in the optimal offline scheduling strategy with makespan $T^*$, the chain $O$ also needs to be executed sequentially. Hence, we have that:
	
	{\small
	\begin{displaymath}
		T^* \geq p_{sum} + \sum\limits_{q \in [Q]}\frac{d_{q}}{\min\{B_{in}^{m^{in}_q}, B_{out}^{m^{out}_q}\}}
	\end{displaymath}}
	In addition, we also have that $T_{OES}$ equals the time for executing whole chain $O$. And in the execution of chain $O$, following Lemma~\ref{lemma:schedule}, the $q$-th flow are transferred with a data rate at least $\min\{B_{in}^{m^{in}_q}/\widehat{\Delta^{m^{in}_q}_{in}}, B_{out}^{m^{out}_q}/\widehat{\Delta^{m^{out}_q}_{out}}\}$.
	Consequently, we have:
	
	{\small
	\begin{equation}
		\label{eqn:path}
		T_{OES} \leq p_{sum} + \sum\limits_{q \in [Q]}\frac{d_{q}}{\min\{B_{in}^{m^{in}_q}/\widehat{\Delta^{m^{in}_q}_{in}}, B_{out}^{m^{out}_q}/\widehat{\Delta^{m^{out}_q}_{out}}\}}
	\end{equation}}
 	Combining the above two inequalities, we have:
	{\small
	\begin{eqnarray*}
		T_{OES} &\leq &  p_{sum} + \sum\limits_{q \in [Q]}\frac{d_{q}}{\min\{B_{in}^{m^{in}_q}/\widehat{\Delta^{m^{in}_q}_{in}}, B_{out}^{m^{out}_q}/\widehat{\Delta^{m^{out}_q}_{out}}\}}
 \\
			& \leq & \Delta(p_{sum} + \sum\limits_{q \in [Q]}\frac{d_{q}}{\min\{B_{in}^{m^{in}_q}, B_{out}^{m^{out}_q}\}}) \\
			& \leq & \Delta \times T^*
	\end{eqnarray*}}

\end{proof}

\section{Proof of Theorem~\ref{theorem:init_placement}}
\label{proof:theorem:init_placement}

\begin{proof}
	Alg.~\ref{alg:init_solution} searches all the tuples $(q_s, q_w, q_{ps}, m)$ through the construction of $\Omega(1)$ to $\Omega(M)$. If a feasible solution exists, it must correspond to one of the tuples, indicating that Alg.~\ref{alg:init_solution} can find the feasible solution.
	
	The construction of $\Omega(1)$ requires $O(\eta_s\eta_w\eta_{ps} R)$ time. Giving $\Omega(m-1)$, for every tuple in it, we can compute all related tuples in $\Omega(m)$ in $O(\eta_s\eta_w\eta_{ps}R)$ time. Since there are at most $O(|J_s||J_w||J_{ps}|)$ tuples in $\Omega(m-1)$, the construction of $\Omega(m)$ giving $\Omega(m-1)$ takes $O(|J_s||J_w||J_{ps}|\eta_s\eta_w\eta_{ps}R)$. Consequently, the time complexity of Alg.~\ref{alg:init_solution} is $O(M|J_s||J_w||J_{ps}|\eta_s\eta_w\eta_{ps}R)$.
\end{proof}

%% file: main.bbl
\begin{thebibliography}{10}
\providecommand{\url}[1]{#1}
\csname url@samestyle\endcsname
\providecommand{\newblock}{\relax}
\providecommand{\bibinfo}[2]{#2}
\providecommand{\BIBentrySTDinterwordspacing}{\spaceskip=0pt\relax}
\providecommand{\BIBentryALTinterwordstretchfactor}{4}
\providecommand{\BIBentryALTinterwordspacing}{\spaceskip=\fontdimen2\font plus
\BIBentryALTinterwordstretchfactor\fontdimen3\font minus
  \fontdimen4\font\relax}
\providecommand{\BIBforeignlanguage}[2]{{%
\expandafter\ifx\csname l@#1\endcsname\relax
\typeout{** WARNING: IEEEtran.bst: No hyphenation pattern has been}%
\typeout{** loaded for the language `#1'. Using the pattern for}%
\typeout{** the default language instead.}%
\else
\language=\csname l@#1\endcsname
\fi
#2}}
\providecommand{\BIBdecl}{\relax}
\BIBdecl

\bibitem{kipf2016semi}
T.~N. Kipf and M.~Welling, ``{Semi-Supervised Classification with Graph
  Convolutional Networks},'' in \emph{Proc. of ICLR}, 2017.

\bibitem{hamilton2017inductive}
W.~L. Hamilton, R.~Ying, and J.~Leskovec, ``{Inductive Representation Learning
  on Large Graphs},'' in \emph{Proc. of NIPS}, 2017.

\bibitem{rusek2020routenet}
K.~Rusek, J.~Su{\'a}rez-Varela, P.~Almasan, P.~Barlet-Ros, and
  A.~Cabellos-Aparicio, ``{RouteNet: Leveraging Graph Neural Networks for
  network modeling and optimization in SDN},'' \emph{IEEE Journal on Selected
  Areas in Communications}, 2020.

\bibitem{li2019encoding}
C.~Li and D.~Goldwasser, ``{Encoding Social Information with Graph
  Convolutional Networks for Political Perspective Detection in News Media},''
  in \emph{Proc. of ACL}, 2019.

\bibitem{duvenaud2015convolutional}
D.~Duvenaud, D.~Maclaurin, J.~Aguilera-Iparraguirre, R.~G{\'o}mez-Bombarelli,
  T.~Hirzel, A.~Aspuru-Guzik, and R.~P. Adams, ``{Convolutional Networks on
  Graphs for Learning Molecular Fingerprints},'' in \emph{Proc. of NIPS}, 2015.

\bibitem{velivckovic2017graph}
P.~Veli{\v{c}}kovi{\'c}, G.~Cucurull, A.~Casanova, A.~Romero, P.~Lio, and
  Y.~Bengio, ``{Graph Attention Networks},'' in \emph{Proc. of ICLR}, 2018.

\bibitem{ErricaPBM20}
F.~Errica, M.~Podda, D.~Bacciu, and A.~Micheli, ``{A Fair Comparison of Graph
  Neural Networks for Graph Classification},'' in \emph{Proc. of ICLR}, 2020.

\bibitem{li2020type}
X.~Li, Y.~Shang, Y.~Cao, Y.~Li, J.~Tan, and Y.~Liu, ``{Type-Aware Anchor Link
  Prediction across Heterogeneous Networks Based on Graph Attention Network},''
  in \emph{Proc. of AAAI}, 2020.

\bibitem{frasconi1998general}
P.~Frasconi, M.~Gori, and A.~Sperduti, ``{A General Framework for Adaptive
  Processing of Data Structures},'' \emph{IEEE Transactions on Neural
  Networks}, 1998.

\bibitem{cao2015grarep}
S.~Cao, W.~Lu, and Q.~Xu, ``{GraRep: Learning Graph Representations with Global
  Structural Information},'' in \emph{Proc.~of ACM CIKM}, 2015.

\bibitem{hu2020open}
W.~Hu, M.~Fey, M.~Zitnik, Y.~Dong, H.~Ren, B.~Liu, M.~Catasta, and J.~Leskovec,
  ``{Open Graph Benchmark: Datasets for Machine Learning on Graphs},'' in
  \emph{Proc.~of NeurIPS}, 2020.

\bibitem{sinha2015overview}
A.~Sinha, Z.~Shen, Y.~Song, H.~Ma, D.~Eide, B.-J. Hsu, and K.~Wang, ``{An
  Overview of Microsoft Academic Service (MAS) and Applications},'' in
  \emph{Proc.~of WWW}, 2015.

\bibitem{zeng2019graphsaint}
H.~Zeng, H.~Zhou, A.~Srivastava, R.~Kannan, and V.~Prasanna, ``{GraphSAINT:
  Graph Sampling Based Inductive Learning Method},'' in \emph{Proc.~of ICLR},
  2020.

\bibitem{chen2018fastgcn}
J.~Chen, T.~Ma, and C.~Xiao, ``{FastGCN: Fast Learning with Graph Convolutional
  Networks via Importance Sampling},'' in \emph{Proc.~of ICLR}, 2018.

\bibitem{zheng2020distdgl}
D.~Zheng, C.~Ma, M.~Wang, J.~Zhou, Q.~Su, X.~Song, Q.~Gan, Z.~Zhang, and
  G.~Karypis, ``{DistDGL: Distributed Graph Neural Network Training for
  Billion-Scale Graphs},'' in \emph{IEEE/ACM Workshop on Irregular
  Applications: Architectures and Algorithms}, 2020.

\bibitem{thorpe2021dorylus}
J.~Thorpe, Y.~Qiao, J.~Eyolfson, S.~Teng, G.~Hu, Z.~Jia, J.~Wei, K.~Vora,
  R.~Netravali, M.~Kim \emph{et~al.}, ``{Dorylus: Affordable, Scalable, and
  Accurate GNN Training with Distributed CPU Servers and Serverless Threads},''
  in \emph{Proc.~of USENIX OSDI}, 2021.

\bibitem{gandhi2021p3}
S.~Gandhi and A.~P. Iyer, ``{P3: Distributed Deep Graph Learning at Scale},''
  in \emph{Proc.~of USENIX OSDI}, 2021.

\bibitem{lin2020pagraph}
Z.~Lin, C.~Li, Y.~Miao, Y.~Liu, and Y.~Xu, ``{PaGraph: Scaling GNN Training on
  Large Graphs via Computation-aware Caching and Partitioning},'' in
  \emph{Proc.~of ACM SoCC}, 2020.

\bibitem{karypis1998fast}
G.~Karypis and V.~Kumar, ``{A Fast and High Quality Multilevel Scheme for
  Partitioning Irregular Graphs},'' \emph{SIAM Journal on Scientific
  Computing}, 1998.

\bibitem{tian2018scheduling}
B.~Tian, C.~Tian, H.~Dai, and B.~Wang, ``{Scheduling Coflows of Multi-stage
  Jobs to Minimize the Total Weighted Job Completion Time},'' in \emph{Proc. of
  IEEE INFOCOM}, 2018.

\bibitem{geyer1992practical}
C.~J. Geyer, ``{Practical Markov Chain Monte Carlo},'' \emph{Statistical
  Science}, 1992.

\bibitem{wang2019deep}
M.~Wang, L.~Yu, D.~Zheng, Q.~Gan, Y.~Gai, Z.~Ye, M.~Li, J.~Zhou, Q.~Huang,
  C.~Ma \emph{et~al.}, ``{Deep Graph Library: Towards Efficient and Scalable
  Deep Learning on Graphs},'' in \emph{ICLR Workshop on Representation Learning
  on Graphs and Manifolds}, 2019.

\bibitem{gilmer2017neural}
J.~Gilmer, S.~S. Schoenholz, P.~F. Riley, O.~Vinyals, and G.~E. Dahl, ``{Neural
  Message Passing for Quantum Chemistry},'' in \emph{Proc.~of ICML}, 2017.

\bibitem{cai2021rethinking}
S.~Cai, L.~Li, J.~Deng, B.~Zhang, Z.-J. Zha, L.~Su, and Q.~Huang, ``{Rethinking
  Graph Neural Architecture Search from Message-passing},'' in \emph{Proc.~of
  IEEE/CVF CVPR}, 2021.

\bibitem{adam2019pytorch}
A.~Paszke, S.~Gross, F.~Massa, A.~Lerer, J.~Bradbury, G.~Chanan, T.~Killeen,
  Z.~Lin, N.~Gimelshein, L.~Antiga, A.~Desmaison, A.~Kopf, E.~Yang, Z.~DeVito,
  M.~Raison, A.~Tejani, S.~Chilamkurthy, B.~Steiner, L.~Fang, J.~Bai, and
  S.~Chintala, ``{PyTorch: An Imperative Style, High-Performance Deep Learning
  Library},'' in \emph{Proc. of NeurIPS}, 2019.

\bibitem{chen2015mxnet}
T.~Chen, M.~Li, Y.~Li, M.~Lin, N.~Wang, M.~Wang, T.~Xiao, B.~Xu, C.~Zhang, and
  Z.~Zhang, ``{MXNet: A Flexible and Efficient Machine Learning Library for
  Heterogeneous Distributed Systems},'' in \emph{NIPS Workshop on Machine
  Learning Systems (LearningSys)}, 2016.

\bibitem{euler}
\BIBentryALTinterwordspacing
(2021) {Euler Graph Library}. [Online]. Available:
  \url{https://github.com/alibaba/euler}
\BIBentrySTDinterwordspacing

\bibitem{abadi2016tensorflow}
M.~Abadi, P.~Barham, J.~Chen, Z.~Chen, A.~Davis, J.~Dean, M.~Devin,
  S.~Ghemawat, G.~Irving, M.~Isard \emph{et~al.}, ``{TensorFlow: A System for
  Large-Scale Machine Learning},'' in \emph{Proc. of USENIX OSDI}, 2016.

\bibitem{zhao2019aligraph}
K.~Zhao, W.~Xiao, B.~Ai, W.~Shen, X.~Zhang, Y.~Li, and W.~Lin, ``{AliGraph: An
  Industrial Graph Neural Network Platform},'' in \emph{Proc.~of SOSP Workshop
  on AI Systems}, 2019.

\bibitem{fey2019fast}
M.~Fey and J.~E. Lenssen, ``{Fast Graph Representation Learning with PyTorch
  Geometric},'' in \emph{Proc.~of ICLR}, 2019.

\bibitem{lambda}
\BIBentryALTinterwordspacing
\emph{{AWS Lambda}}, 2021. [Online]. Available:
  \url{https://aws.amazon.com/lambda}
\BIBentrySTDinterwordspacing

\bibitem{ma2019neugraph}
L.~Ma, Z.~Yang, Y.~Miao, J.~Xue, M.~Wu, L.~Zhou, and Y.~Dai, ``{NeuGraph:
  Parallel Deep Neural Network Computation on Large Graphs},'' in
  \emph{Proc.~of USENIX ATC}, 2019.

\bibitem{jia2020improving}
Z.~Jia, S.~Lin, M.~Gao, M.~Zaharia, and A.~Aiken, ``{Improving the Accuracy,
  Scalability, and Performance of Graph Neural Networks with ROC},'' in
  \emph{Proc.~of MLSys}, 2020.

\bibitem{wang2021flexgraph}
L.~Wang, Q.~Yin, C.~Tian, J.~Yang, R.~Chen, W.~Yu, Z.~Yao, and J.~Zhou,
  ``{FlexGraph: A Flexible and Efficient Distributed Framework for GNN
  Training},'' in \emph{Proc.~of ACM EuroSys}, 2021.

\bibitem{chiang2019cluster}
W.-L. Chiang, X.~Liu, S.~Si, Y.~Li, S.~Bengio, and C.-J. Hsieh, ``{Cluster-GCN:
  An Efficient Algorithm for Training Deep and Large Graph Convolutional
  Networks},'' in \emph{Proc. of ACM KDD}, 2019.

\bibitem{cai2021dgcl}
Z.~Cai, X.~Yan, Y.~Wu, K.~Ma, J.~Cheng, and F.~Yu, ``{DGCL: An Efficient
  Communication Library for Distributed GNN Training},'' in \emph{Proc.~of ACM
  EuroSys}, 2021.

\bibitem{zhang2017poseidon}
H.~Zhang, Z.~Zheng, S.~Xu, W.~Dai, Q.~Ho, X.~Liang, Z.~Hu, J.~Wei, P.~Xie, and
  E.~P. Xing, ``{Poseidon: An Efficient Communication Architecture for
  Distributed Deep Learning on GPU Clusters},'' in \emph{Proc.~of USENIX ATC},
  2017.

\bibitem{Jayarajan2019p3}
A.~Jayarajan, J.~Wei, G.~Gibson, A.~Fedorova, and G.~Pekhimenko,
  ``{Priority-Based Parameter Propagation for Distributed DNN Training},'' in
  \emph{Proc.~of Systems and Machine Learning (SysML)}, 2019.

\bibitem{shi2021mg}
S.~Shi, X.~Chu, and B.~Li, ``{MG-WFBP: Merging Gradients Wisely for Efficient
  Communication in Distributed Deep Learning},'' \emph{IEEE Transactions on
  Parallel and Distributed Systems}, 2021.

\bibitem{bao2019deep}
Y.~Bao, Y.~Peng, and C.~Wu, ``{Deep Learning-based Job Placement in Distributed
  Machine Learning Clusters},'' in \emph{Proc.~of IEEE INFOCOM}, 2019.

\bibitem{mirhoseini2017device}
A.~Mirhoseini, H.~Pham, Q.~V. Le, B.~Steiner, R.~Larsen, Y.~Zhou, N.~Kumar,
  M.~Norouzi, S.~Bengio, and J.~Dean, ``{Device Placement Optimization with
  Reinforcement Learning},'' in \emph{Proc.~of ICML}, 2017.

\bibitem{wang2020geryon}
S.~Wang, D.~Li, and J.~Geng, ``{Geryon: Accelerating Distributed CNN Training
  by Network-level Flow Scheduling},'' in \emph{Proc.~of IEEE INFOCOM}, 2020.

\bibitem{park2020hetpipe}
J.~H. Park, G.~Yun, M.~Y. Chang, N.~T. Nguyen, S.~Lee, J.~Choi, S.~H. Noh, and
  Y.-r. Choi, ``{HetPipe: Enabling Large DNN Training on (Whimpy) Heterogeneous
  GPU Clusters through Integration of Pipelined Model Parallelism and Data
  Parallelism},'' in \emph{Proc.~of USENIX ATC}, 2020.

\bibitem{yi2020optimizing}
X.~Yi, S.~Zhang, Z.~Luo, G.~Long, L.~Diao, C.~Wu, Z.~Zheng, J.~Yang, and
  W.~Lin, ``{Optimizing Distributed Training Deployment in Heterogeneous GPU
  Clusters},'' in \emph{Proc. of ACM CoNEXT}, 2020.

\bibitem{brown2006traffic}
M.~A. Brown, ``{Traffic Control HOWTO},'' \emph{Guide to IP Layer Network},
  2006.

\bibitem{goodfellow2016deep}
I.~Goodfellow, Y.~Bengio, and A.~Courville, \emph{{Deep Learning}}.\hskip 1em
  plus 0.5em minus 0.4em\relax MIT Press, 2016.

\bibitem{ogbnpapers100msage}
\BIBentryALTinterwordspacing
\emph{GraphSAGE\_res\_incep}, 2021. [Online]. Available:
  \url{https://github.com/mengyangniu/ogbn-papers100m-sage}
\BIBentrySTDinterwordspacing

\bibitem{tan2019joint}
H.~Tan, S.~H.-C. Jiang, Y.~Li, X.-Y. Li, C.~Zhang, Z.~Han, and F.~C.~M. Lau,
  ``{Joint Online Coflow Routing and Scheduling in Data Center Networks},''
  \emph{IEEE/ACM Transactions on Networking}, 2019.

\end{thebibliography}
